%% file: complete_E.tex
\documentclass[12pt,english,aip, reprint]{revtex4-1}
\usepackage[T1]{fontenc}
\usepackage[latin9]{inputenc}
\usepackage{refstyle}
\usepackage{amsmath}
\usepackage{graphicx}

\makeatletter

\AtBeginDocument{\providecommand\secref[1]{\ref{sec:#1}}}
\AtBeginDocument{\providecommand\eqref[1]{\ref{eq:#1}}}
\AtBeginDocument{\providecommand\figref[1]{\ref{fig:#1}}}
\AtBeginDocument{\providecommand\tabref[1]{\ref{tab:#1}}}
\AtBeginDocument{\providecommand\subsecref[1]{\ref{subsec:#1}}}
\RS@ifundefined{subsecref}
  {\newref{subsec}{name = \RSsectxt}}
  {}
\RS@ifundefined{thmref}
  {\def\RSthmtxt{theorem~}\newref{thm}{name = \RSthmtxt}}
  {}
\RS@ifundefined{lemref}
  {\def\RSlemtxt{lemma~}\newref{lem}{name = \RSlemtxt}}
  {}

\usepackage[version=3]{mhchem}
\usepackage{url}
\usepackage{notes2bib}

\@ifundefined{showcaptionsetup}{}{%
 \PassOptionsToPackage{caption=false}{subfig}}
\usepackage{subfig}
\makeatother

\usepackage{babel}
\begin{document}
\let\stepref=\relax
\newref{step}{
	name = step~,
	names = steps~,
	Name = Step~,
	Names = Steps~
}

\newcommand*\ie{{\em i.e.}}
\newcommand*\eg{{\em e.g.}}
\newcommand*\etal{{\em et al.}}
\newcommand*\citeref[1]{ref. \citenum{#1}}
\newcommand*\citerefs[1]{refs. \citenum{#1}} 

\newcommand*\Erkale{{\sc Erkale}}
\newcommand*\LibXC{{\sc LibXC}}
\newcommand*\PySCF{{\sc PySCF}}
\newcommand*\PsiFour{{\sc Psi4}}
\title{Polarized Gaussian basis sets from one-electron ions}
\author{Susi Lehtola}
\affiliation{Department of Chemistry, University of Helsinki, P.O. Box 55 (A. I. Virtasen
aukio 1), FI-00014 University of Helsinki, Finland}
\email{susi.lehtola@alumni.helsinki.fi}

\begin{abstract}
We demonstrate that basis sets suitable for electronic structure calculations
can be obtained from simple accuracy considerations for the hydrogenic
one-electron ions $Y^{(Y-1)+}$ for $Y\in[1,Z]$, necessitating no
self-consistent field calculations at all. It is shown that even-tempered
basis sets with parameters from the commonly-used universal Gaussian
basis set (UGBS) {[}E. V. R. de Castro and F. E. Jorge, J. Chem. Phys.
108, 5225 (1998){]} reproduce non-relativistic spin-restricted spherical
Hartree--Fock total energies from fully numerical calculations to
better accuracy than UGBS, which is shown to exhibit huge errors for
some elements, \eg{} 0.19 $E_{h}$ for \ce{Th+} and 0.13 $E_{h}$
for Lu as it has been parametrized for a single atomic configuration.
Having shown the feasibility of the one-electron approach, partially
energy-optimized basis sets are formed for all atoms in the periodic
table, $1\leq Z\leq118$, by optimizing the even-tempered parameters
for $Z^{(Z-1)+}$. As the hydrogenic Gaussian basis sets suggested
in the present work are built strictly from first principles, also
polarization shells can be obtained in the same fashion in contrast
to previous approaches. The accuracy of the polarized basis sets is
demonstrated by calculations on a small set of molecules by comparison
to fully numerical reference values, which show that chemical accuracy
can be reached even for challenging cases like \ce{SF6}. The present
approach is straightforward to extend to relativistic calculations,
and could facilitate studies beyond the established periodic table.
\end{abstract}
\maketitle

\section{Introduction\label{sec:Introduction}}

Basis sets are the keystones of quantum chemistry, as they are used
to define the allowed degrees of freedom for the one-particle states
of the electrons, that is, the molecular orbitals $\psi_{i}$ as
\begin{equation}
\psi_{i}(\boldsymbol{r})=\sum_{\mu}C_{\mu i}\chi_{\mu}(\boldsymbol{r}).\label{eq:basexp}
\end{equation}
\Eqref{basexp} leads to a discretization of the electronic structure
problem, which can then be solved on a computer. Although several
kinds of basis sets can be adopted for molecular electronic structure
calculations (see \eg{} \citeref{Lehtola2019c} for an overview),
a linear combination of atomic orbitals
\begin{equation}
\chi_{nlm}(\boldsymbol{r})=R_{nl}(r)Y_{l}^{m}(\hat{\boldsymbol{r}})\label{eq:AO}
\end{equation}
is the traditional choice, as it usually affords an excellent level
of cancellation of systematic errors in the study of energy differences.
Due to the facility they afford by analytic integral evaluation, Gaussian
basis sets 
\begin{equation}
R_{nl}^{\text{GTO}}=r^{l}\sum_{k}c_{k}^{nl}e^{-\alpha_{k}^{nl}r^{2}}\label{eq:gto}
\end{equation}
have been the overwhelmingly dominant type of atomic orbital used
in computational chemistry for a long time, and a plethora of Gaussian
basis sets tailored for various purposes have been published in the
literature; see \eg{} \citerefs{Davidson1986, Jensen2013b, Hill2013}
for further details. 

While typical applications to modeling chemical reactions use compact
Gaussian basis sets that have been tightly optimized and carefully
constructed for this specific purpose, they do not always fulfill
all the needs of a computational chemist. Few sets cover all of the
periodic table in a uniform manner, and even fewer can be used for
all-electron calculations, the rest employing effective core potentials;
the situation is, however, improving as systematic basis sets covering
(almost) the whole periodic table are becoming available.\citep{Roos2004,Pollak2017,Zobel2020} 

Large uncontracted basis sets are often necessary for accurate studies
of atoms and small molecules with Gaussian-basis electronic structure
programs, but they are often not readily available. If a large enough
basis set is used, it has accrued sufficient variational freedom to
become universal, meaning it can be used for all atoms.\citep{Silver1978,Silver1978a}
Because of this, universal basis sets are often useful for benchmarking
purposes, as they are typically available for all elements, which
is often not the case with commonly-used, carefully optimized basis
sets. Fully numerical basis sets such as those used in \citerefs{Lehtola2019a, Lehtola2019b, Lehtola2019e}
are an excellent proof of the concept of universal basis sets: a few
hundred radial basis functions suffice to reproduce Hartree--Fock
and density functional energies beyond microhartree accuracy for all
atoms in the periodic table (H--Og) at the non-relativistic level
of theory.\citep{Lehtola2019a,Lehtola2019e} 

The universal Gaussian basis set\citep{DeCastro1998} (UGBS) is likely
the best-known example of universal Gaussian basis sets,\citep{Mohallem1987,DaCosta1987,DaSilva1989,Jorge1997,Jorge1997a}
and at present is the all-electron basis that supports the most atoms
according to the Basis Set Exchange.\citep{Pritchard2019} The UGBS
has been parametrized for H--Lr excluding Pa--Np, Cm, and Bk,\citep{DeCastro1998}
and employs a common set of exponents on for all angular momenta that
range from $O(10^{-2})$ to $O(10^{8})$, depending on the atom. Because
of its accuracy and wide availability, the UGBS of \citeref{DeCastro1998}
has been used in a wide variety of studies including (but not limited
to) atomic charges;\citep{Manz2010} optimized effective,\citep{Ryabinkin2013,Kollmar2014}
model Kohn--Sham,\citep{Kohut2014} modified Slater,\citep{Gaiduk2015}
Fermi,\citep{Ospadov2018} and exact exchange-correlation potentials;\citep{Kanungo2019}
steps in the Kohn--Sham potential,\citep{Hodgson2017} visualization
of atomic sizes,\citep{Ospadov2018a} simplified relativistic calculations,\citep{Wang2000a}
the basis set convergence of spin-spin coupling constants,\citep{Deng2006}
the characterization of density functionals,\citep{Haunschild2010a,Carmona-Espindola2015}
semi-numerical implementations of relativistic exact exchange,\citep{Maier2019a}
strongly repulsive interatomic potentials,\citep{Lehtola2020a} and
as a starting point for new energy-optimized basis sets;\citep{Pantazis2009,Pantazis2011,Pantazis2012,Pantazis2014}
we refer the reader to the literature for more details. In addition
to the UGBS of \citeref{DeCastro1998} discussed above and the likewise
commonly-used universal basis set for Rydberg states,\citep{Kaufmann1989}
there are a number of other universal Gaussian basis sets that have
been reported in the literature but do not appear to have become as
well known. These include the relativistic universal Gaussian basis
sets\citep{Malli1993,Malli1994,Haiduke2005} as well as many others;
we again refer the reader to the literature as a sufficiently thorough
overview cannot be presented here.

However, universal Gaussian basis sets are problematic for molecular
calculations. Due to their size, universal basis sets like UGBS may
cause severe issues with linear dependencies in molecular calculations;
however, a routine solution to this problem has been presented recently.\citep{Lehtola2019f,Lehtola2020a}
Next, since the basis sets are typically obtained from atomic multiconfigurational
Hartree--Fock calculations for a single state, they are not guaranteed
to be accurate for other atomic states. Moreover, polarization functions
that are needed for accuracy in molecular calculations with Hartree--Fock
or density functional theory\citep{Hohenberg1964,Kohn1965} (DFT)
and/or correlation functionals that are necessary for post-Hartree--Fock
calculations have to be construed in some other manner. Although it
is possible in principle to employ a common set of exponents for the
occupied and unoccupied shells (as is done in \citeref{Deng2006},
for example), the resulting basis sets are untractably large for most
applications, as unnecessarily many polarization functions are produced.

In the present work, we will show that polarized basis sets can be
constructed via completeness-optimization\citep{Manninen2006,Lehtola2015}
on hydrogenic atoms; a similar approach was recently used successfully
for determining fully numerical basis sets for diatomic molecules.\citep{Lehtola2019b}
The method developed in the present work allows rapid generation of
novel basis sets with pre-estimated accuracy, without the need for
costly electronic structure calculations involving iterative techniques.
We will describe the method in detail in \secref{Method}, and demonstrate
it for non-relativistic spin-restricted Hartree--Fock calculations
on neutral atoms and their cations employing the even-tempered parameters
from UGBS in \secref{Results}. Having showed that the universal basis
sets thus obtained are more accurate than UGBS, we then form partially
energy-optimized basis sets, demonstrate their accuracy for atomic
calculations, as well as showcase their performance on a series of
molecular calculations. The work is summarized and conclusions are
presented in \secref{Summary-and-Conclusions}. Atomic units are used
throughout the text.

\section{Method\label{sec:Method}}

\subsection{Hydrogenic wave functions\label{subsec:Hydrogenic-wave-functions}}

The idea for using hydrogenic ions as a proxy for determining if the
basis set is complete enough arises from our recent work on initial
guesses for electronic structure theory.\citep{Lehtola2019} The atomic
orbitals for any atom can be in principle obtained from a scalar radial
potential $V^{\text{eff}}(r)$, which features a screened nuclear
charge. Rewriting the radial potential, which is easy to extract from
density functional calculations,\citep{Lehtola2019,Lehtola2019e}
in terms of the Coulomb potential of an effective nuclear charge $V^{\text{eff}}(r)=-Z^{\text{eff}}(r)/r$
shows that near the core, the electrons experience the full nuclear
charge, while far away the exact potential has a $-1/r$ behavior
but approximate density functional potentials decay exponentially.\citep{Lehtola2019}
At intermediate ranges, the nuclear charge falls somewhere in-between
the full nuclear charge and the unit nuclear charge.

As we now know the form of the potential the electrons are moving
in, an exceedingly simple recipe for building Gaussian basis sets
can be postulated: if the basis set can represent the ground states
of all one-electron ions from neutral hydrogen to the extreme cation
$Z^{(Z-1)+}$, it should likewise do a good job at describing the
electronic structure of the atom with a full set of interacting electrons.

A similar conclusion can also be reached for polarization functions:
they, too, experience the same (unknown) scalar potential $V^{\text{eff}}(r)$
as the other electrons. What makes polarization functions different
from occupied orbitals is the higher kinetic energy arising from the
$l(l+1)/r^{2}$ term. Since orbital energy denominators arise both
in post-Hartree--Fock methods and self-consistent field perturbation
theory,\citep{Maschio2020} the $l(l+1)/r^{2}$ term means that regardless
of the electronic structure method, tight exponents become less and
less important as $l$ grows---and this effect can already be captured
by the study of the hydrogenic ions.

However, the \emph{indirect} effect of the $l(l+1)/r^{2}$ term is
not captured by the hydrogenic ions: the effective charge is smaller
further away from the nucleus.\citep{Lehtola2019} For instance, the
$1s$, $2s$ and $2p$ orbitals are a fraction of the size of the
$3d$ orbital in Kr, meaning that the $3d$ orbital does not experience
the full nuclear charge. This screening of the charge could be employed
to limit the range of $Z$ in increasing $l$. However, we will consider
the full $Z$ for all $l$ values for the present purpose of proof
of principle, since part of the orbitals always tunnel through to
small $r$ where the charge is less-screened; an adequate representation
of tight functions with large $l$ may also be necessary for post-Hartree--Fock
approaches with core correlation.

As a large number of hydrogenic calculations are required for the
optimization, a specialized implementation is used to solve the hydrogenic
problems. It is straightforward to derive the elements of the overlap
$\boldsymbol{S}$, nuclear attraction $\boldsymbol{V}^{\text{nuc}}$
and kinetic energy $\boldsymbol{T}$ matrices in the basis defined
in \eqref{gto} by use of standard techniques (see \eg{} \citeref{Lehtola2019a}).
The matrices are diagonal in $l$ and $m$ and carry no $m$ dependence,\citep{Lehtola2019a}
meaning that the exponents can be determined independently in each
$l$ block. The expressions within each block turn out to be exceedingly
simple
\begin{align}
S_{ij} & =\frac{1}{2}\Gamma\left(l+\frac{3}{2}\right)\left(\alpha_{i}+\alpha_{j}\right)^{-l-3/2}\label{eq:overlap}\\
V_{ij}^{\text{nuc}} & =-\frac{1}{2}\Gamma(l+1)\left(\alpha_{i}+\alpha_{j}\right)^{-l-1}\label{eq:nuclear}\\
T_{ij} & =\left(l+\frac{3}{2}\right)\Gamma\left(l+\frac{3}{2}\right)\frac{\alpha_{i}\alpha_{j}}{\left(\alpha_{i}+\alpha_{j}\right)^{l+5/2}}\label{eq:kinetic}
\end{align}
Given a set of exponents $\{\alpha_{i}\}$, the hydrogenic energy
can be computed by solving the generalized eigenvalue problem 
\begin{equation}
\left(\boldsymbol{T}+Z\boldsymbol{V}^{\text{nuc}}\right)\boldsymbol{C}=\boldsymbol{S}\boldsymbol{C}\boldsymbol{E}\label{eq:eval}
\end{equation}
where $\boldsymbol{C}$ and $\boldsymbol{E}$ are the matrix of orbital
coefficients and diagonal matrix of orbital energies $e_{i}$; the
lowest $e_{i}$ yielding the energy of the hydrogenic ground state
of $Z^{(Z-1)+}$. \Eqref{eval} is solved using a canonically orthonormalized
basis set,\citep{Lowdin1956} in which a $10^{-7}$ threshold is used
to eliminate any linear dependencies in the basis.

\subsection{Completeness-optimization\label{subsec:Completeness-optimization}}

It is well known that full optimization of basis sets with $N\gg1$
exponents turns out to yield sequences that resemble a geometric one
\begin{equation}
\alpha_{i}=\alpha_{0}\beta^{i}\text{ for }i\in0,1,\dots N-1\label{eq:et}
\end{equation}
in the middle part; see \eg{} \citerefs{Jensen1999} and \citeref{Bakken2004}
for an illustration on energy-optimization on the hydrogen atom, and
\citeref{Lehtola2015} for completeness-optimized primitives. In a
full optimization of the primitives,\citep{Petersson2003} the outermost
energy-optimized exponents move out compared to the sequence of \eqref{et},\citep{Jensen2000,Bakken2004}
whereas in completeness-optimization the outermost exponents move
\emph{in}.\citep{Lehtola2015}

The primitives arising from \eqref{et} are known as even-tempered.\citep{Raffenetti1973a}
Even-tempered expansions are interesting for their simplicity. Expressing
$N\gg1$ exponents in terms of two parameters $\alpha_{0}$ and $\beta$
makes it simple to generate large expansions that are appreciably
close to the fully optimal ones, and the sets are trivial to augment
with further tight and diffuse functions for basis set convergence
studies. Most importantly, even-tempered exponents span the Hilbert
space evenly,\citep{Cherkes2009} and become complete when $\alpha_{0}\to0$,
$\beta\to1$, and $N\to\infty$;\citep{Feller1979} accurate molecular
properties can be achieved by approaching this limit. Moreover, $\alpha_{0}$
and $\beta$ optimized for individual atoms are close to optimal also
in a molecular environment if a large enough basis set is used.\citep{Feller1979} 

As was already mentioned in the Introduction, the present approach
works by completeness-optimizing the basis set. The completeness of
a basis set can be quantified as its ability to represent a given
test function with parameter $\alpha$, as measured by the norm of
its projection onto the basis set\citep{Chong1995}
\begin{equation}
Y(\alpha)=\sum_{\mu\nu}\langle\alpha|\mu\rangle\langle\mu|\nu\rangle^{-1}\langle\nu|\alpha\rangle\label{eq:cpl}
\end{equation}
from which $0\leq Y(\alpha)\leq1$; the test function for Gaussian
basis sets is typically chosen as a primitive Gaussian and the parameter
$\alpha$ as the test function's exponent. 

A completeness-optimized basis set\citep{Manninen2006} maximizes
the completeness profile $Y(\alpha)$ for some range of exponents
$\alpha\in[\alpha_{\min},\alpha_{\max}]$, where the limits $\alpha_{\min}$
and $\alpha_{\max}$ are determined by trial and error for the property
in question.\citep{Manninen2006} Although \eqref{cpl} suggests a
way to optimize the primitives in the basis for given values $\alpha_{\min}$
and $\alpha_{\max}$ as was already hinted above (see details in \citeref{Lehtola2015}),
since the idea in completeness-optimization is to expand the limits
$\alpha_{\min}$ and $\alpha_{\max}$ until the property no longer
changes, we believe completeness-optimization of the primitives is
unnecessary and that a simple even-tempered expansion should suffice.
(The $\beta$ parameter, however, could be fixed based on completeness
arguments; see the Appendix.)

The procedure for the completeness-optimization of the proposed hydrogenic
basis sets proceeds as follows. First, the values for $\alpha_{0}$
and $\beta$ of the even-tempered sequence are fixed. For the value
of $\alpha_{0}$ and $\beta$ we will use the UGBS values $\alpha_{0}^{\text{UGBS}}=0.02000046$
and $\beta^{\text{UGBS}}=1.958150$.\citep{DeCastro1998} Next, we
allow exponents smaller than $\alpha_{0}$ to be produced by letting
the index $i$ have negative values in \eqref{et}. As scaling $\alpha_{0}\to\alpha_{0}\beta^{j}$
with integer $j$ is tantamount to relabeling the indices $i\to i+j$
in \eqref{et}, $\alpha_{0}$ only matters modulo $\beta$ and can
be restricted without loss of generality to $\alpha_{0}\in[1,\beta)$.
The value used for $\alpha_{0}$ makes no difference in the completeness
argument: if a large enough set of exponents is used, any choice of
$\alpha_{0}$ should yield the same answer at the end. (Note that
in the complete basis set limit $\beta\to1$ as was discussed above,\citep{Feller1979}
and $\alpha_{0}$ indeed becomes irrelevant.) To prove that the choice
for $\alpha_{0}$ is unimportant, in the following we will also examine
the case of maximally different exponents obtained by choosing $\alpha_{0}\to\alpha_{0}\sqrt{\beta}$
in addition to studying the UGBS value of $\alpha_{0}$.

Having chosen the permitted grid of exponents $\alpha_{i}$ according
to \eqref{et}, the optimal single exponent $\alpha_{i}$ that minimizes
the energy of the single-electron ion $Z^{(Z-1)+}$ (\eqref{eval})
is found. Then, steeper exponents $i+1,i+2,\dots$ as well as more
diffuse exponents $i-1,i-2,\dots$ are added one by one until the
change in the hydrogenic energy converges to a threshold $\epsilon(Z)$;
this defines the basis set for the ion $Z^{(Z-1)+}$ as a range of
exponents $i\in\left[i_{\min}\left(Z^{(Z-1)+}\right),i_{\max}\left(Z^{(Z-1)+}\right)\right]$.
The basis set for an element can then be acquired as simply as taking
$i_{\min}=\min_{Z}i_{\min}\left(Z^{(Z-1)+}\right)$ and $i_{\max}=\max_{Z}i_{\max}\left(Z^{(Z-1)+}\right)$,
as this should satisfy the requirement that all one-electron ions
from H to $Z^{(Z-1)+}$ be reproduced with the specified accuracy.

Because the exact hydrogenic ground-state energy in each $l$ channel
\begin{equation}
e_{i}=-Z^{2}/[2(l+1)^{2}]\label{eq:Ehyd}
\end{equation}
has steep scaling in $Z$, we define the threshold used in the profile
extension as
\begin{equation}
\epsilon(Z)=\frac{Z^{2}}{\lg\beta}\epsilon,\label{eq:thresh}
\end{equation}
which essentially means that the hydrogenic energy should be reproduced
to a relative accuracy of $\epsilon$. The $\lg\beta$ factor has
been added into \eqref{thresh} to normalize the threshold to unit
profile increment. Covering large ranges of exponent space becomes
slow for small $\beta$ values; the normalization with $\lg\beta$
should make the input threshold $\epsilon$ transferable accross $\beta$
values.

The composition of the unpolarized basis set was chosen as follows:
$s$ shell only for H and He, $s$ and $p$ shells for Li--Ar; $s$,
$p$, and $d$ shells for K--Xe; and $s$, $p$, $d$, and $f$ shells
for Cs--Og. Because the above procedure for choosing the primitives
does not depend on the role of the shell, occupied, polarization and
correlation shells are obtained in an equal fashion. We have parametrized
basis sets with up to 3 polarization/correlation shells that range
up to $i$ functions in the present work.

\section{Results\label{sec:Results}}

\subsection{Universal basis sets}

The \PySCF{} program\citep{Sun2018} is used for non-relativistic
spin-restricted Hartree--Fock (NRSRHF) calculations employing spherically
averaged densities. Even though NRSRHF theory is not accurate for
chemistry, it is sufficient for the present purposes of probing whether
the basis sets are capable of qualitative electronic structure calculations.
(Here, we note that UGBS has likewise been parametrized for non-relativistic
calculations.\citep{DeCastro1998}) The Gaussian-basis results from
\PySCF{} are compared with reference values computed with the fully
numerical approach of \citeref{Lehtola2019e}; the truncation error
of the Gaussian basis set is extracted by substracting the fully numerical
reference values from the Gaussian-basis results.

\begin{figure}
\begin{centering}
\subfloat[$\epsilon=10^{-3}$]{\begin{centering}
\includegraphics[width=0.5\textwidth]{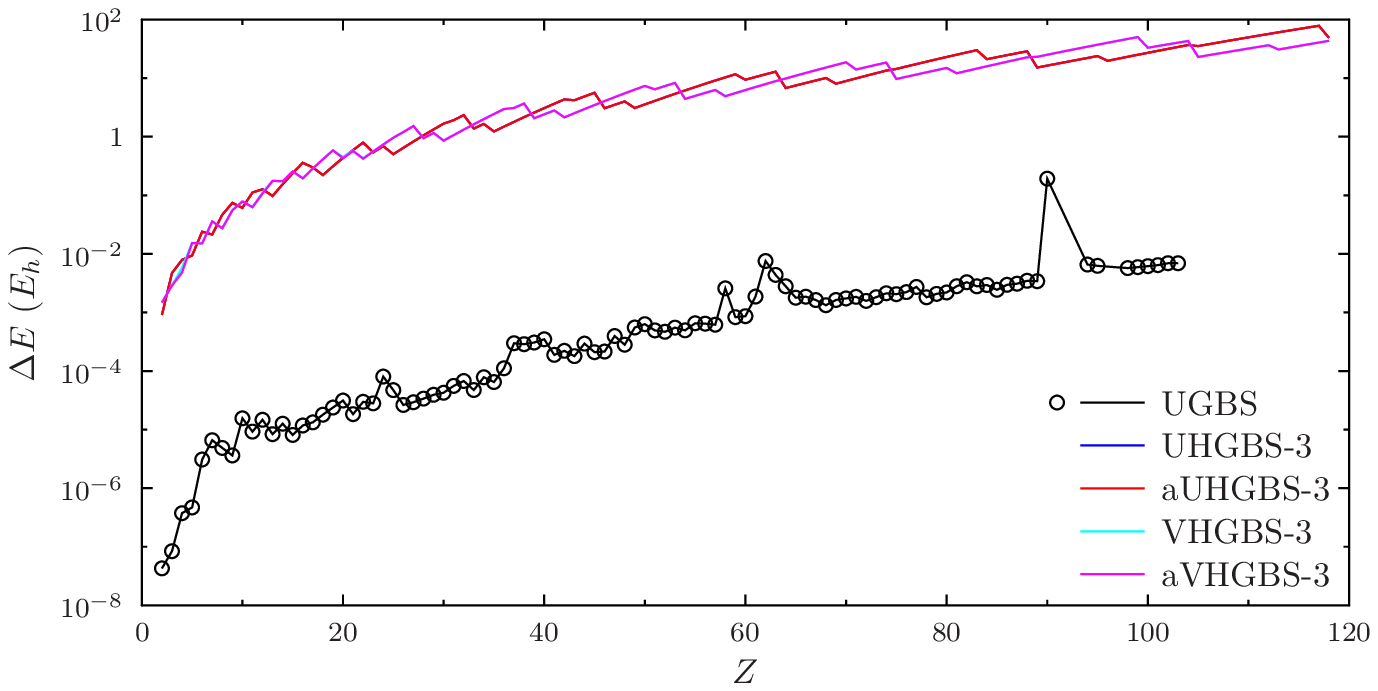}
\par\end{centering}

}
\par\end{centering}
\begin{centering}
\subfloat[$\epsilon=10^{-5}$]{\begin{centering}
\includegraphics[width=0.5\textwidth]{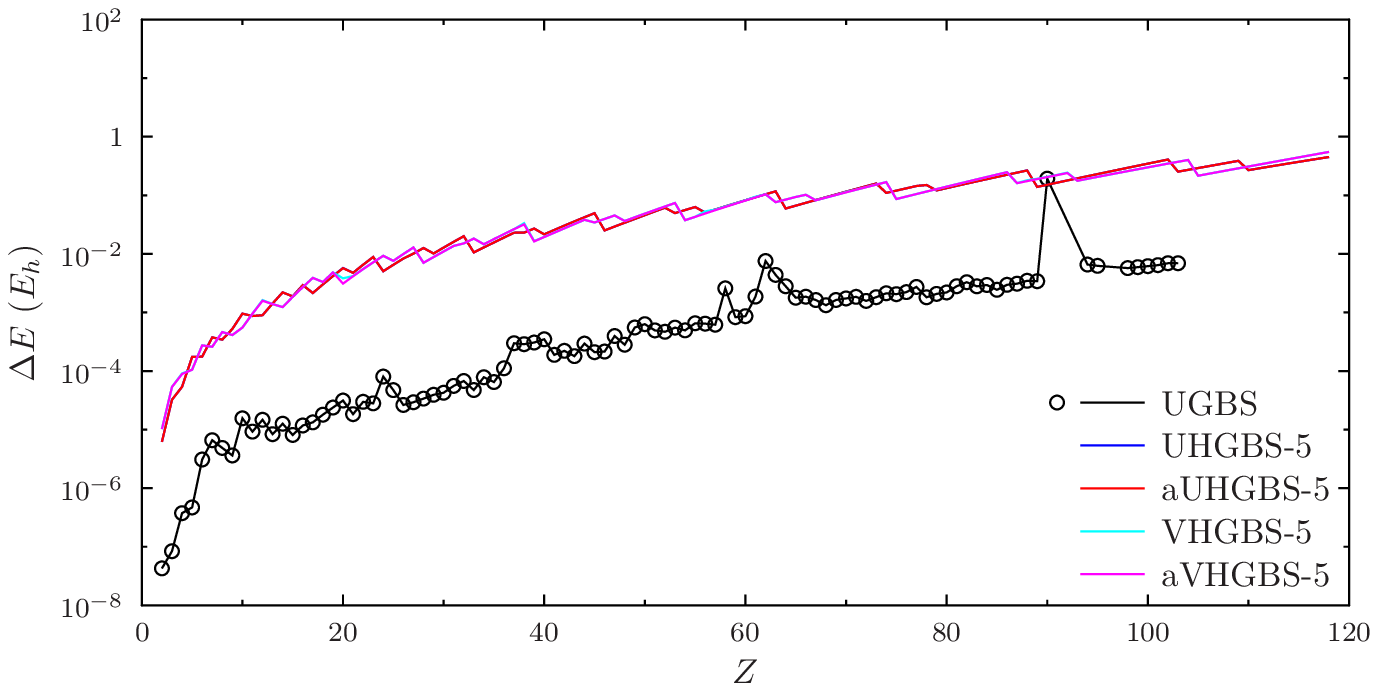}
\par\end{centering}
}
\par\end{centering}
\begin{centering}
\subfloat[$\epsilon=10^{-7}$]{\begin{centering}
\includegraphics[width=0.5\textwidth]{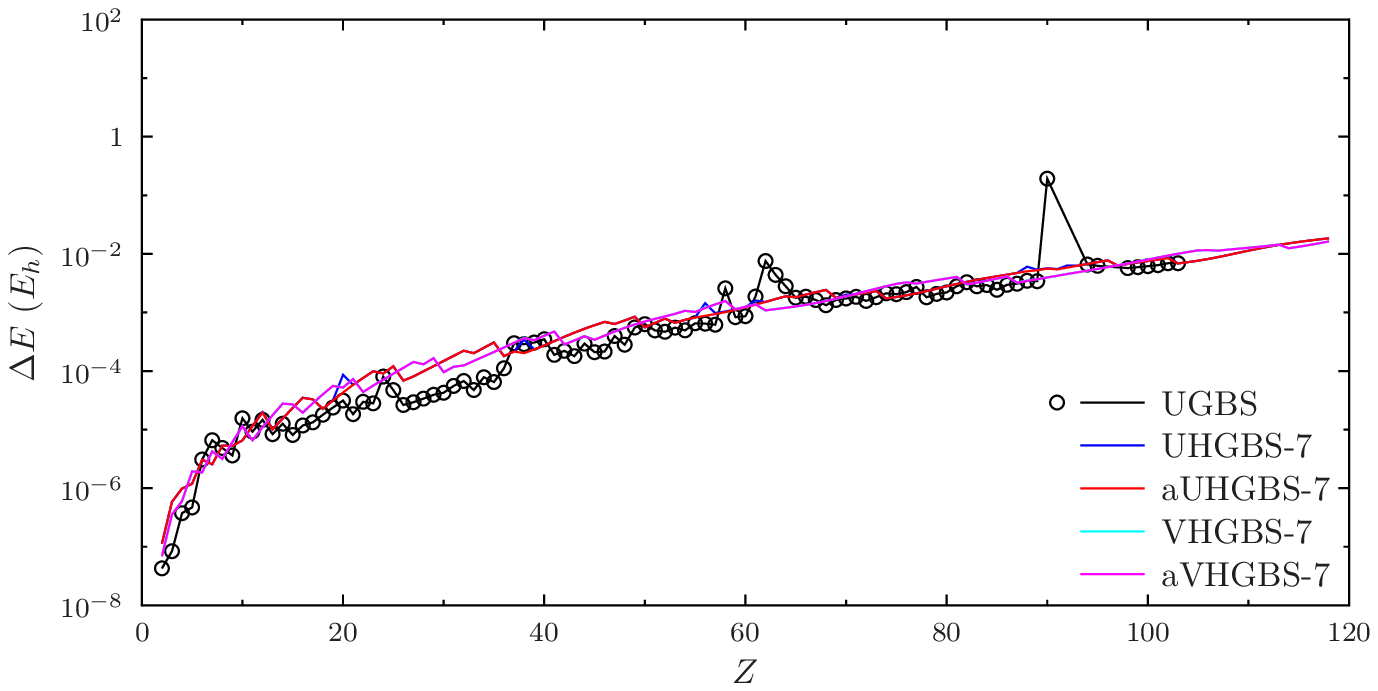}
\par\end{centering}
}
\par\end{centering}
\begin{centering}
\subfloat[$\epsilon=10^{-9}$]{\begin{centering}
\includegraphics[width=0.5\textwidth]{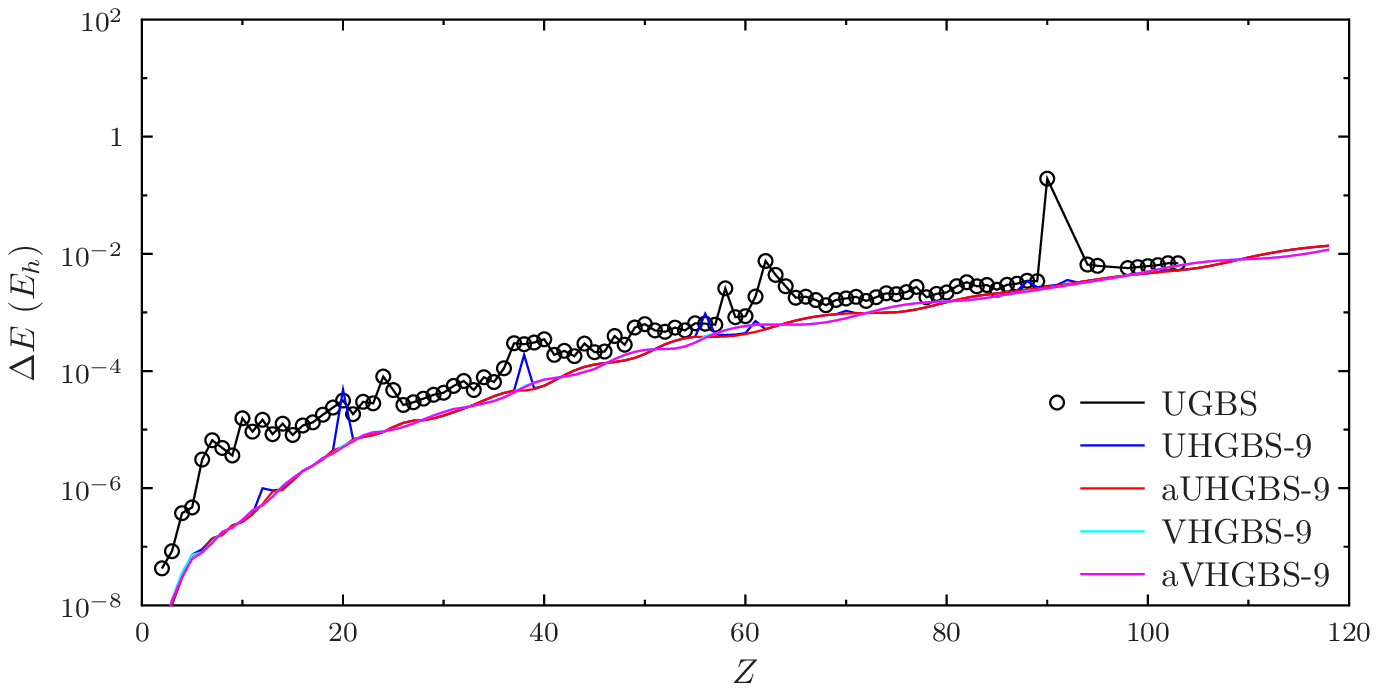}
\par\end{centering}
}
\par\end{centering}
\caption{Truncation error in the total energy of atomic cations at varying
levels of the basis set expansion threshold $\epsilon$ with $\beta=\beta^{\text{UGBS}}\approx1.96$.\label{fig:cations}}
\end{figure}

The accuracy of four novel basis sets at varying thresholds $\epsilon$
is demonstrated in \figref{cations} for the cations $Z^{+}$, whose
ground-state configurations and energies are shown in \tabref{HF-gs}.
Hydrogenic basis sets parametrized to an error of $\epsilon=10^{-n}$
are denoted with the $-n$ suffix, and are available as part of the
supplementary material. The universal hydrogenic gaussian basis set
(UHGBS) employs the UGBS values for the parameters $\alpha_{0}$ and
$\beta$. Results are also shown for the choice $\alpha_{0}\to\alpha_{0}\sqrt{\beta}$
that leads upon completeness-optimization to the VHGBS basis set,
V being the next letter in the alphabet after U. Energy-optimized
augmented basis sets are typically formed by the study of anions.
As anions are typically less bound than neutral atoms, we form augmented
versions of the UHGBS and VHGBS basis sets, aUHGBS and aVHGBS, respectively,
by considering the fictitious $Z=0.5$ one-electron ion. For comparison,
a suitable copy of the UGBS basis set was obtained from the Basis
Set Exchange.\citep{Pritchard2019} 

The atomic energies have large errors with basis sets formed with
low thresholds $\epsilon$ and show piecewise character in $Z$. However,
an universal improvement in accuracy is obtained by tightening the
threshold. A reliable reproduction of the energies of cationic atoms
is achieved for $\epsilon=10^{-9}$, for which the basis set error
behaves smoothly in $Z$. As is clearly seen in \figref{cations},
the new UGBS-style basis sets predict significantly lower energies
than the literature UGBS basis set that behaves less smoothly in $Z$
and shows significant errors for some atoms, \eg{} $1.92\times10^{-1}E_{h}$
for \ce{Th+} and $7.56\times10^{-3}E_{h}$ for \ce{Sm+}. In comparison,
the largest error for the UHGBS basis with $\epsilon=10^{-9}$ in
the range of $Z$ covered by UGBS (see Introduction) is $5.26\times10^{-3}E_{h}$
for $Z=102$.

Small kinks can still be seen for $\epsilon=9$ for the UHGBS basis
set in \figref{cations} at $Z=12$, $Z=20$, $Z=38$, $Z=56$, $Z=61$,
$Z=88$, and $Z=92$; that is for \ce{Mg+}, \ce{Ca+}, \ce{Sr+},
\ce{Ba+}, \ce{Pm+}, \ce{Ra+} and \ce{U+}. All of the kinks go
away upon augmentation, leaving only the two smooth interveawing curves
corresponding to the different choices for $\alpha_{0}$. This confirms
that the minor problems in the UHGBS and VHGBS basis sets have to
do with insufficient diffuse functions for some atoms. Mg, Ca, Sr,
Ba, and Ra are alkaline metals that are well-known to have diffuse
electronic structure; the NRSRHF calculation for their cations also
result in a diffuse $ns$ orbital, which is apparently not moving
in a $-1/r$ potential assumed in the construction of the basis set. 

\begin{figure*}
\subfloat[UGBS]{\begin{centering}
\includegraphics[width=0.8\columnwidth]{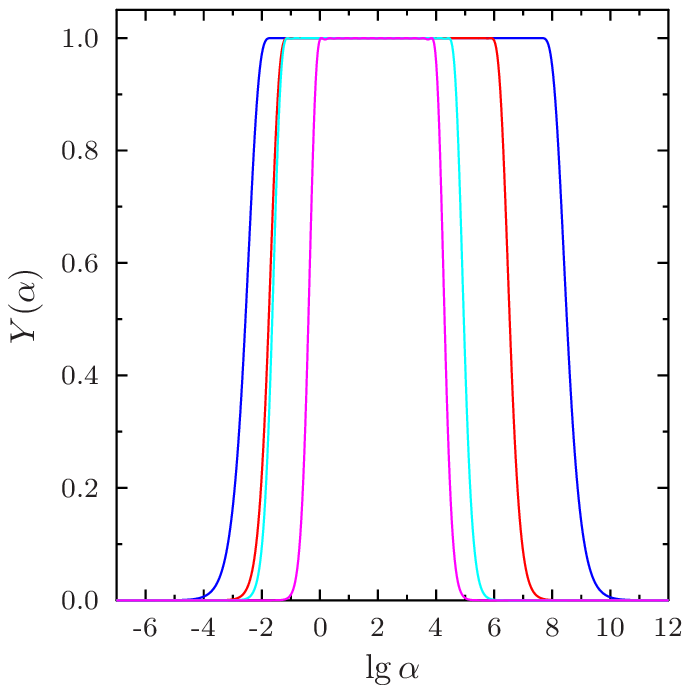}
\par\end{centering}
}\subfloat[UHGBS-9]{\begin{centering}
\includegraphics[width=0.8\columnwidth]{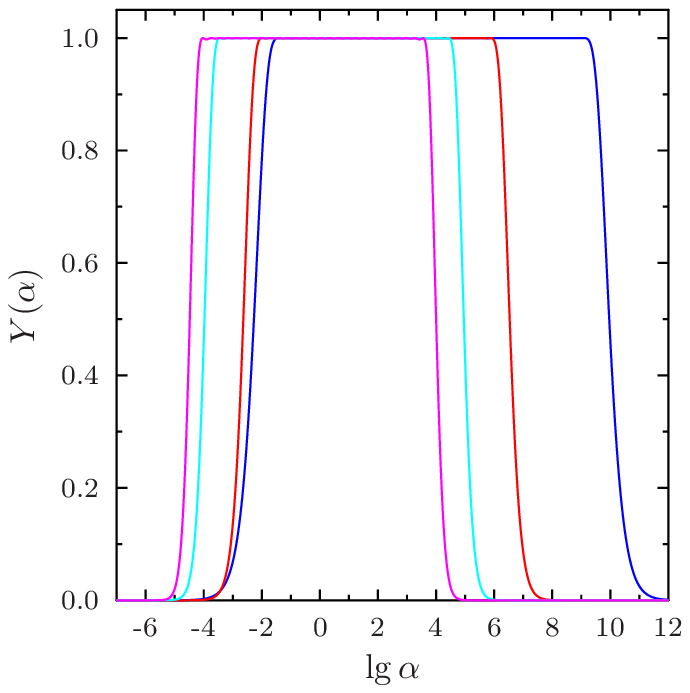}
\par\end{centering}
}

\caption{The completeness profiles for the UGBS and UHGBS-9 thorium basis sets
for the $s$ (blue), $p$ (red), $d$ (cyan), and $f$ (magenta) shells.\label{fig:Th-profile}}
\end{figure*}

To investigate the large truncation error for \ce{Th+} in the UGBS
basis set, completeness profiles for thorium for the UGBS and UHBGS-9
basis sets were computed with \Erkale{}\citep{erkale,Lehtola2012}
and are shown in \figref{Th-profile}. The largest $p$, $d$, and
$f$ exponents are similar in UGBS and UHGBS-9---anecdotally confirming
the validity of the present scheme---while the latter has more tight
$s$ functions, and also considerably more diffuse $d$ and $f$ functions
which likely arise from the neglect of screening far away from the
nucleus. The NRSRHF ground state for \ce{Th+} is $7s^{2}5f^{1}$
(\tabref{HF-gs}) while the UGBS basis set has been parametrized for
the $7s^{2}6d^{2}$ configuration; comparing the completeness profiles
in \figref{Th-profile} suggests that the $f$ shell isn't sufficiently
well sampled by UGBS. 

\begin{figure}
\begin{centering}
\subfloat[$\epsilon=10^{-3}$]{\begin{centering}
\includegraphics[width=0.5\textwidth]{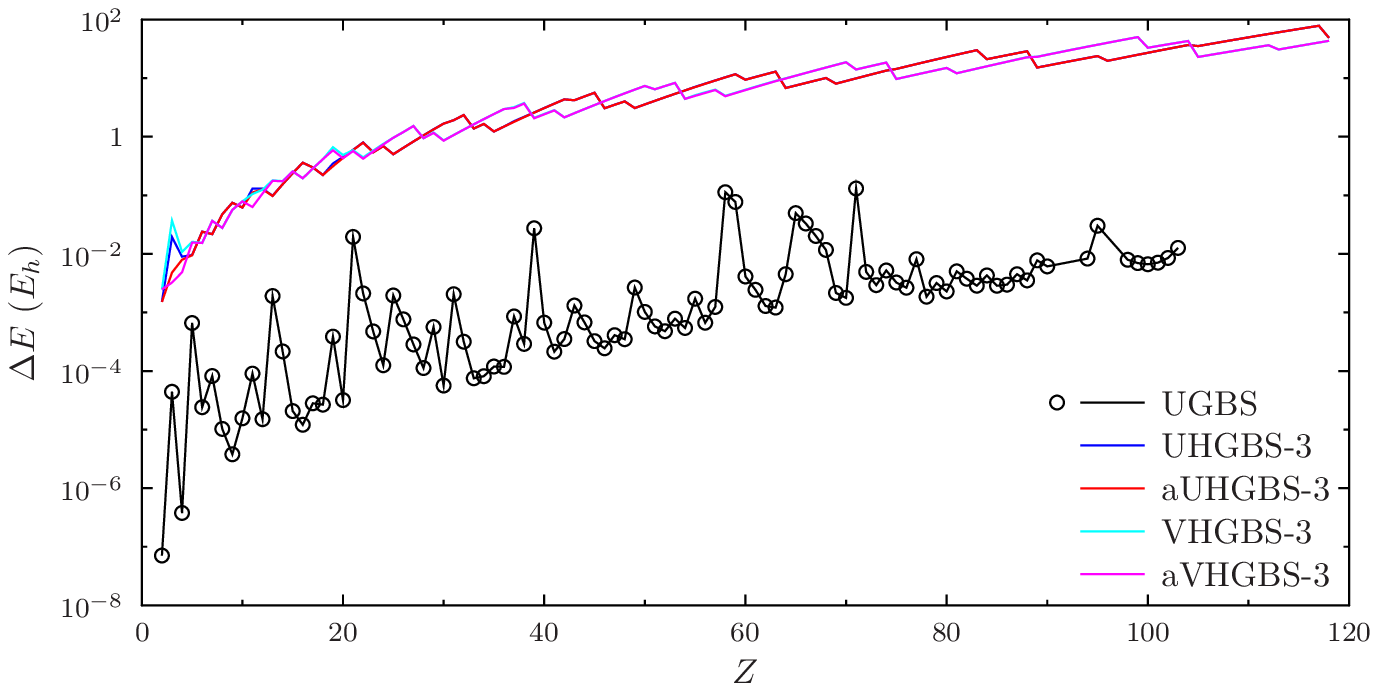}
\par\end{centering}
}
\par\end{centering}
\begin{centering}
\subfloat[$\epsilon=10^{-5}$]{\begin{centering}
\includegraphics[width=0.5\textwidth]{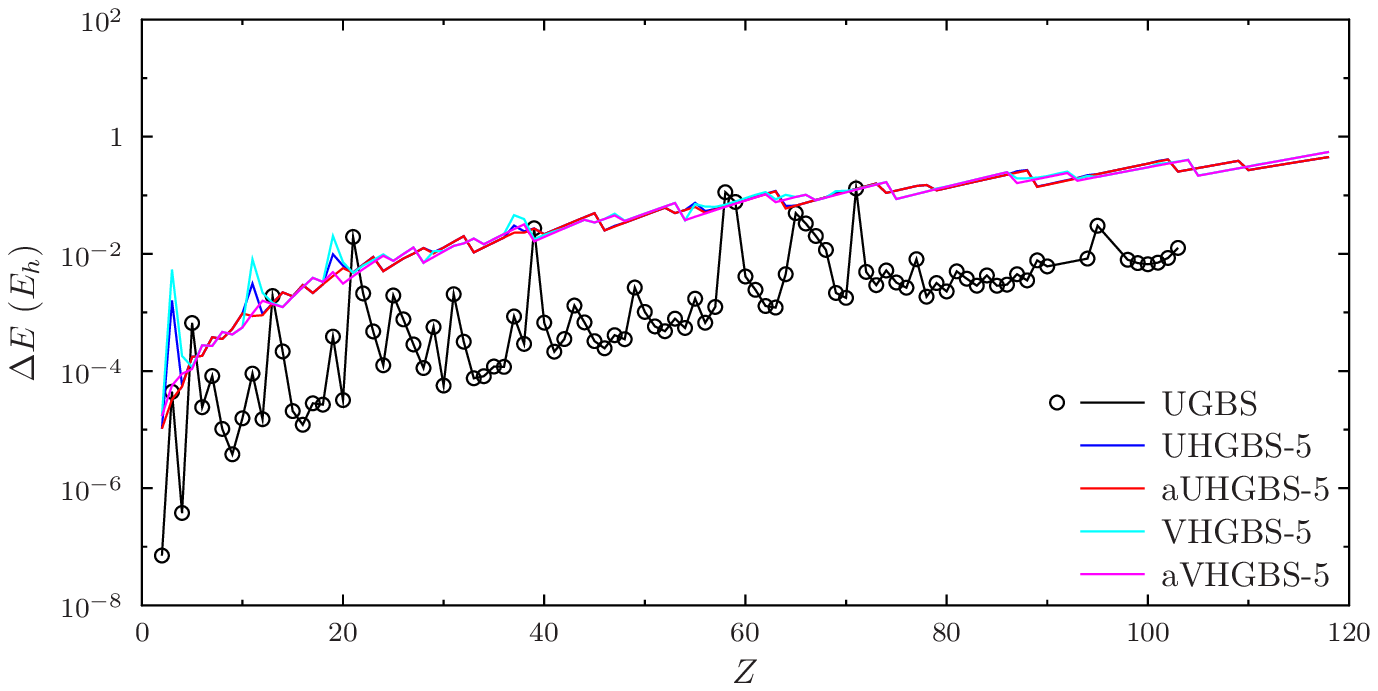}
\par\end{centering}
}
\par\end{centering}
\begin{centering}
\subfloat[$\epsilon=10^{-7}$]{\begin{centering}
\includegraphics[width=0.5\textwidth]{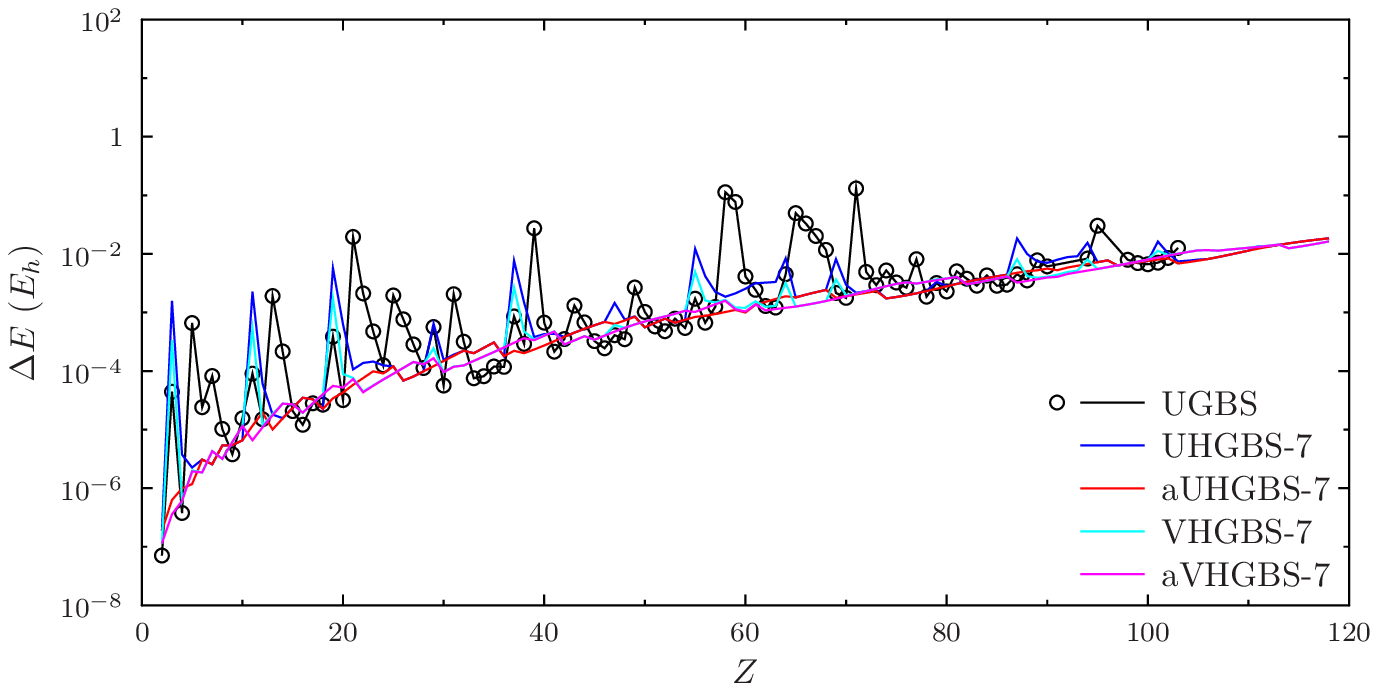}
\par\end{centering}
}
\par\end{centering}
\begin{centering}
\subfloat[$\epsilon=10^{-9}$]{\begin{centering}
\includegraphics[width=0.5\textwidth]{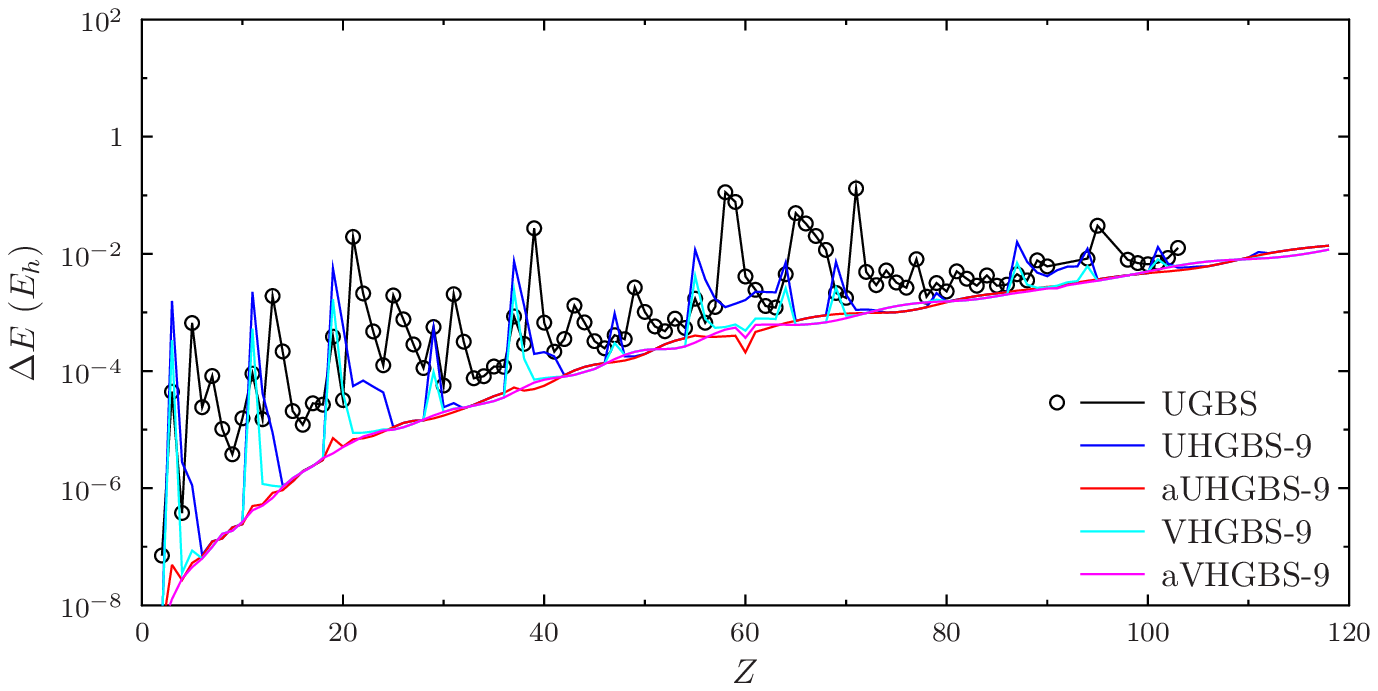}
\par\end{centering}
}
\par\end{centering}
\caption{Truncation error in the total energy of neutral atoms at varying levels
of the basis set expansion threshold $\epsilon$ with $\beta=\beta^{\text{UGBS}}\approx1.96$.\label{fig:neutral}}
\end{figure}

Analogous results for the neutral atoms $Z$ are shown in \figref{neutral},
with configurations and reference energies from \citeref{Lehtola2019e}.
The alkali atoms Li, Na, K, Rb, Cs, and Fr stand out in \figref{neutral}
like their chemical analogues, the alkaline cations, did in in \figref{cations};
the alkali peaks in \figref{neutral} also have a shoulder from the
alkaline atoms. In addition, Sc--Cr stand out in the UHGBS results,
as do Cu, Ag, and Au; the lanthanoid sequence Cs--Gd and Tm, and
the actinoid sequence Fr--U and Md. As with the cations, also these
errors go down significantly when an augmented basis set is used,
suggesting that the outermost electrons are moving in a potential
that is weaker than $-1/r$.

In contrast to the overall smooth behavior of the hydrogenic basis
sets even in the present case of the neutral atoms, the UGBS basis
set shows several large errors: \eg{} $1.31\times10^{-1}E_{h}$ for
Lu, $1.13\times10^{-1}E_{h}$ for Ce, $7.70\times10^{-2}E_{h}$ for
Pr, $4.92\times10^{-2}E_{h}$ for Tb, $3.30\times10^{-2}E_{h}$ for
Dy, $3.02\times10^{-2}$ for Am, $2.74\times10^{-2}E_{h}$ for Y,
$2.02\times10^{-2}E_{h}$ for Ho, and $1.95\times10^{-2}E_{h}$ for
Sc. In contrast, the largest truncation error for UHGBS-9 is $1.59\times10^{-2}E_{h}$
for Fr which is the heaviest alkali atom.

UGBS fails differently for the neutral atoms and for the cations,
because the NRSRHF configurations are sometimes pronouncedly dissimilar
for the two charge states. Neutral thorium is not a problem for UGBS,
because the NRSRHF ground state for Th is\citep{Lehtola2019e} $7s^{2}6d^{2}$---the
same configuration for which UGBS has been optimized\citep{DeCastro1998}---while
the $7s^{2}5f^{1}$ configuration was used for \ce{Th+}, as was discussed
above. In the case of Lu, the NRSRHF ground state is $4f^{14}6s^{2}6p^{1}$
while UGBS has been parametrized for $4f^{14}6s^{2}5d^{1}$. Also
the other large errors of UGBS appear to follow the same pattern:
$6s^{2}4f^{2}5d^{1}$ (low-lying NRSFRHF excited state, see \citeref{Lehtola2019e})
vs $6s^{2}4f^{3}$ for Pr, $4f^{11}$ vs $6s^{2}4f^{9}$ for Tb, $4f^{12}$
vs $6s^{2}4f^{10}$ for Dy, $5f^{9}$ vs $7s^{2}5f^{7}$ for Am, $5s^{2}5p^{1}$
vs $5s^{2}4d^{1}$ for Y, $4f^{13}$ vs $6s^{2}4f^{11}$ for Ho, and
$4s^{2}4p^{1}$ vs $4s^{2}3d^{1}$ for Sc.

Results for a larger choice of $\beta=\left(\beta^{\text{UGBS}}\right)^{1.5}\approx2.74$
are shown in the supporting information. While a basis set limit is
again routinely obtained by the completeness-optimization, yielding
smooth curves for small $\epsilon$, the truncation errors become
significant (tens of $E_{h}$) for the superheavy atoms.

\begin{table*}
\small
\input{HF+.tex}

\caption{Non-relativistic spin-restricted HF configurations with spherically
averaged densities for all cations in the periodic table, employing
the same methodology as for the neutral atoms in \citeref{Lehtola2019e}.\label{tab:HF-gs}}
\end{table*}

\subsection{Energy-optimized sets}

\begin{figure}
\begin{centering}
\subfloat[$\epsilon=10^{-3}$]{\begin{centering}
\includegraphics[width=0.5\textwidth]{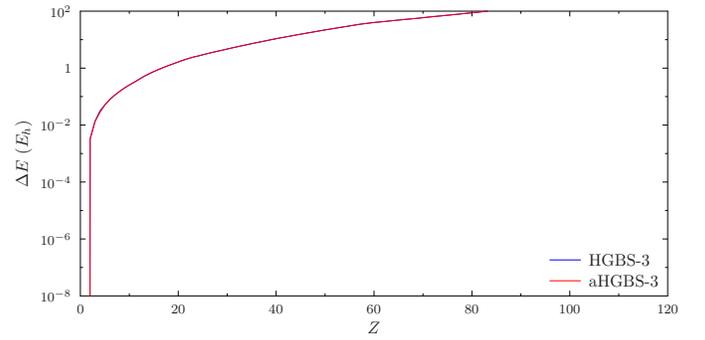}
\par\end{centering}
}
\par\end{centering}
\begin{centering}
\subfloat[$\epsilon=10^{-5}$]{\begin{centering}
\includegraphics[width=0.5\textwidth]{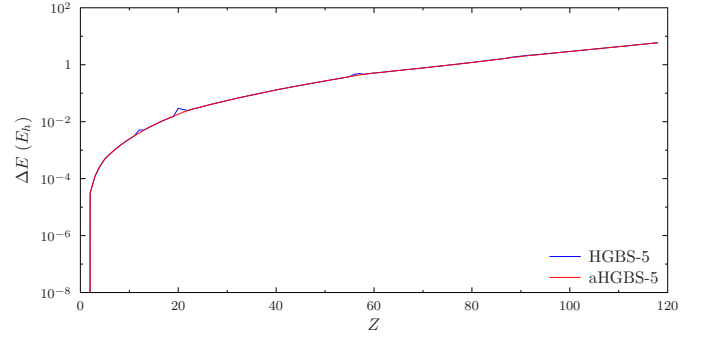}
\par\end{centering}
}
\par\end{centering}
\begin{centering}
\subfloat[$\epsilon=10^{-7}$]{\begin{centering}
\includegraphics[width=0.5\textwidth]{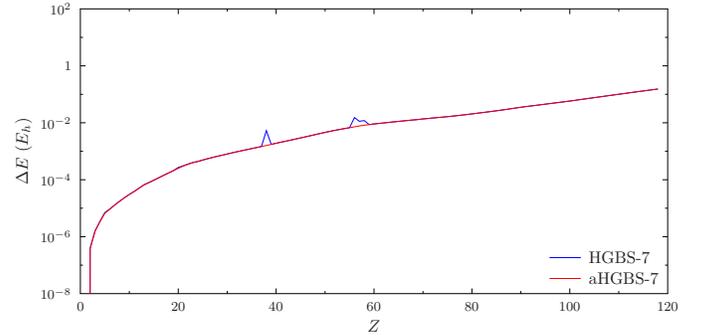}
\par\end{centering}
}
\par\end{centering}
\begin{centering}
\subfloat[$\epsilon=10^{-9}$]{\begin{centering}
\includegraphics[width=0.5\textwidth]{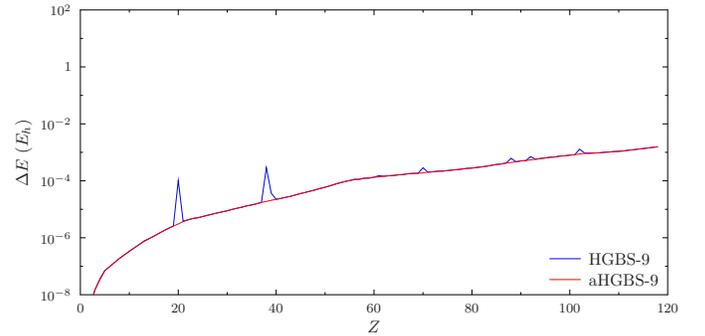}
\par\end{centering}
}
\par\end{centering}
\caption{Truncation error in the total energy of atomic cations with energy-optimized
sets for various thresholds $\epsilon$.\label{fig:cations-opt}}
\end{figure}

\begin{figure}
\begin{centering}
\subfloat[$\epsilon=10^{-3}$]{\begin{centering}
\includegraphics[width=0.5\textwidth]{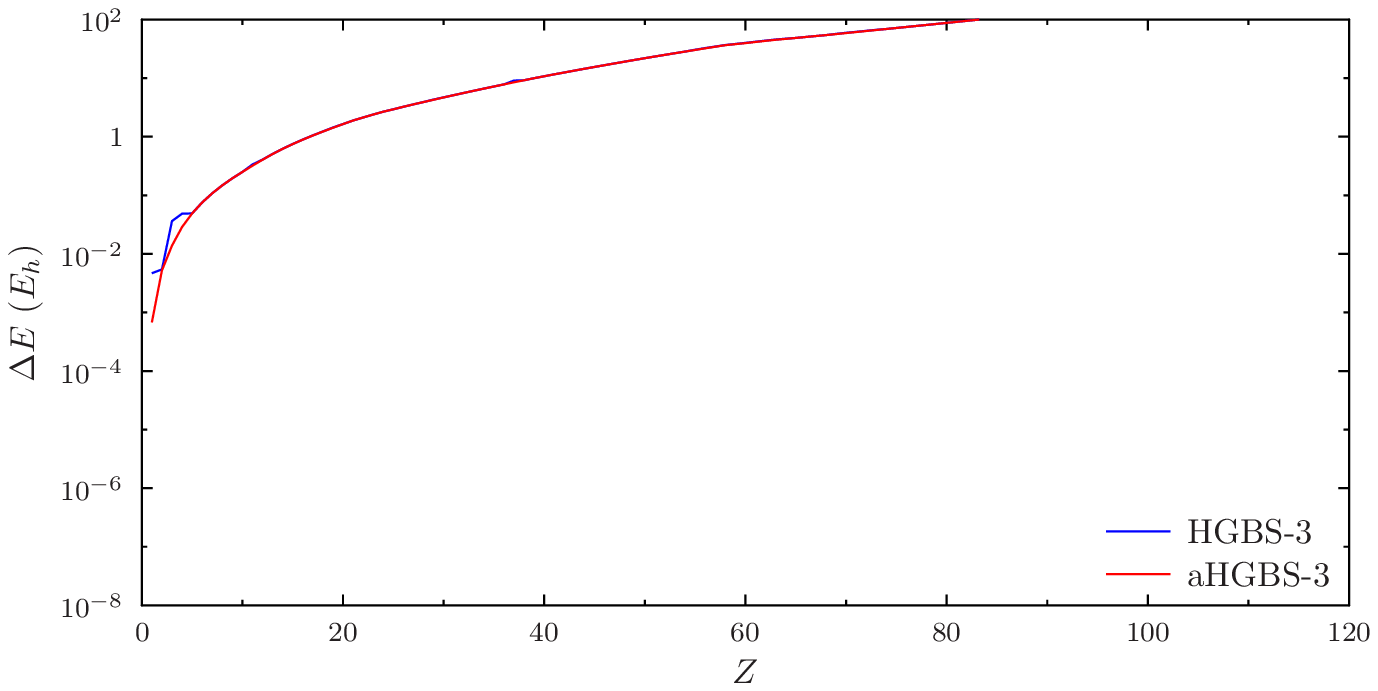}
\par\end{centering}
}
\par\end{centering}
\begin{centering}
\subfloat[$\epsilon=10^{-5}$]{\begin{centering}
\includegraphics[width=0.5\textwidth]{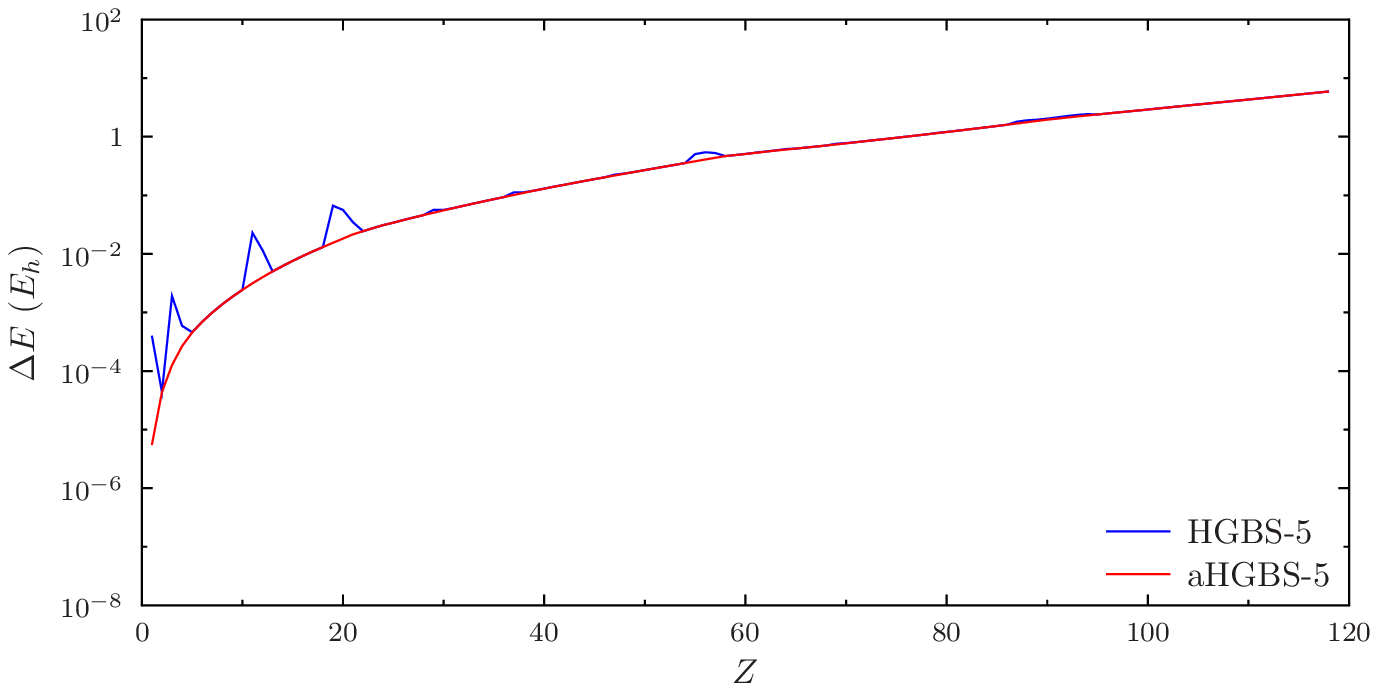}
\par\end{centering}
}
\par\end{centering}
\begin{centering}
\subfloat[$\epsilon=10^{-7}$]{\begin{centering}
\includegraphics[width=0.5\textwidth]{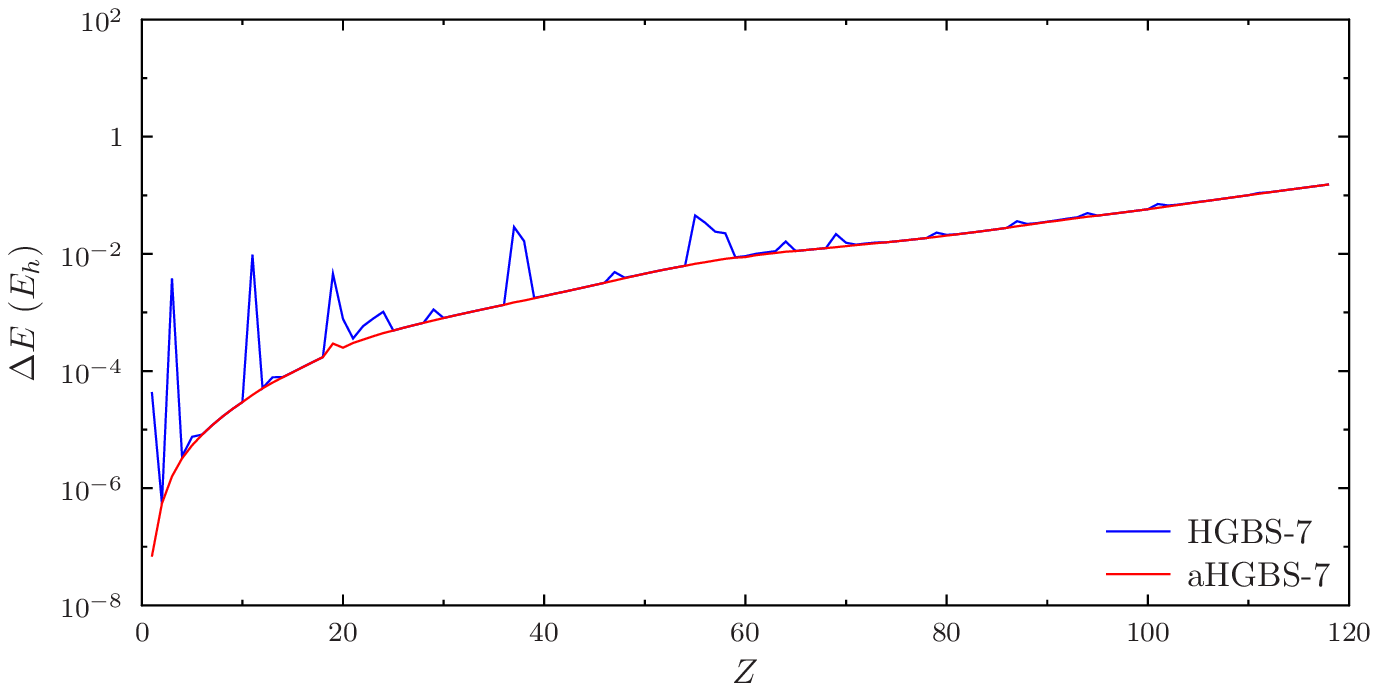}
\par\end{centering}
}
\par\end{centering}
\begin{centering}
\subfloat[$\epsilon=10^{-9}$]{\begin{centering}
\includegraphics[width=0.5\textwidth]{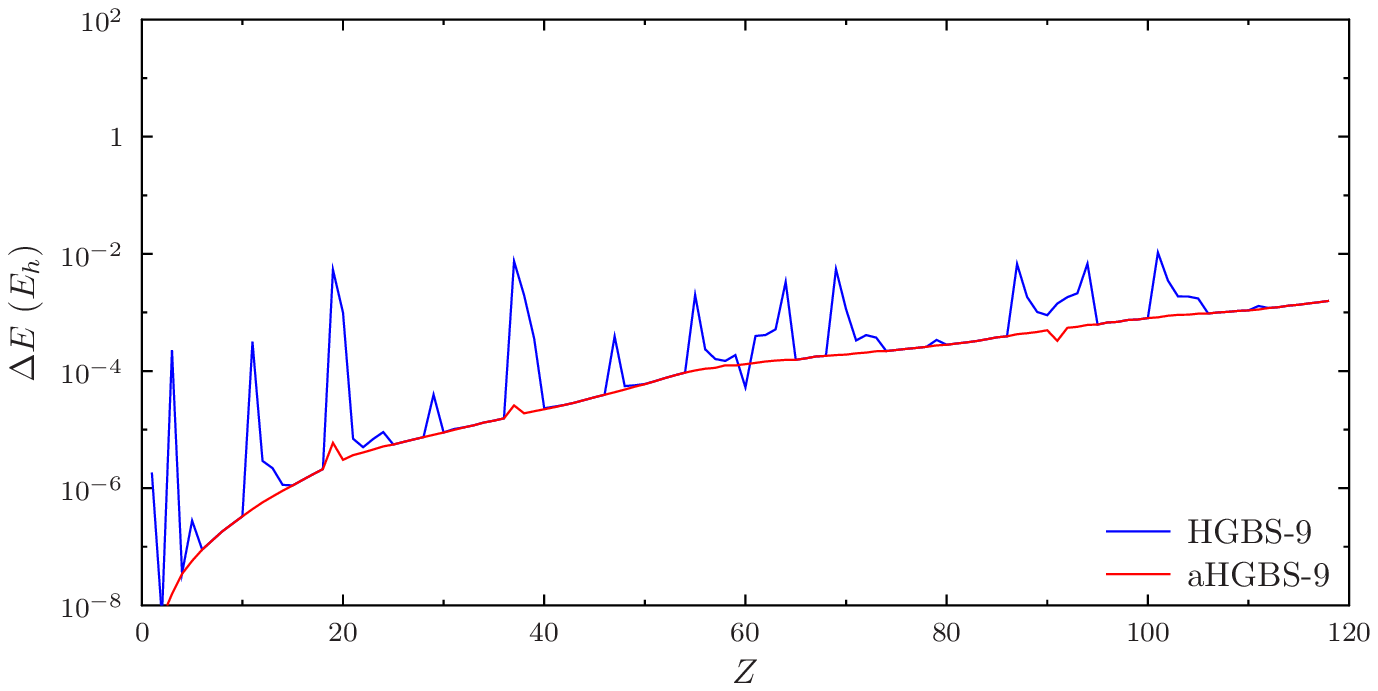}
\par\end{centering}
}
\par\end{centering}
\caption{Truncation error in the total energy of neutral atoms with energy-optimized
sets for various thresholds $\epsilon$.\label{fig:neutral-opt}}
\end{figure}

We have shown that accurate universal basis sets can be formed from
one-electron ions. However, the use of a universal set of exponents
on all atoms is suboptimal. As was seen above, heavy atoms require
smaller values of $\beta$ than light atoms do, indicating that a
smaller basis set with constant accuracy over $Z$ can be obtained
by making $\alpha_{0}$ and $\beta$ free parameters in the formation
of the basis set. Having shown that the one-electron ions offer a
reasonable way to form basis sets, instead of fixing $\alpha_{0}$
and $\beta$ beforehand we can obtain them from minimization of the
energy of $Z^{(Z-1)+}$; otherwise, the basis sets are formed as before.
Although the parameters depend on the length of the expansion ($\alpha_{0}\to0$
and $\beta\to1$ for the complete basis set limit\citep{Feller1979}),
the natural choice is to optimize a sequence of $\alpha_{0}$ and
$\beta$ for an increasing number of functions, and stop when the
addition of an exponent results in an incremental change of the energy
smaller than $Z^{2}\epsilon$. (The $\lg\beta$ parameter in \eqref{thresh}
can be dropped, as $\beta$ is now explicitly optimized for the shell.)

The accuracy of the basis sets that have been partially optimized
in the aforementioned way for NRSRHF calculations on atomic cations
is shown in \figref{cations-opt}. Due to the optimization per atom,
the sets are not universal: the even-tempered parameters $\alpha_{0}$
and $\beta$ are determined separately for every element $Z$ and
every angular momentum $l$. We will refer to the partially optimized
basis sets simply as hydrogenic gaussian basis sets, HGBS. The results
for the HGBS basis sets are again excellent: only a few outliers that
disappear upon augmentation are observed. The HGBS-9 basis set (corresponding
to $\epsilon=10^{-9}$) has truncation errors only in the order of
m$E_{h}$ for superheavy atoms, compared to tens of m$E_{h}$ for
the universal UHGBS-9 basis.

The corresponding accuracy on the neutral atoms is shown in \figref{neutral-opt}.
As with the UGBS-style basis sets, more outliers are seen for the
neutral atoms than for the cations, as the added outermost electron
is less tightly bound, leading to a more diffuse electronic structure.
Also these outliers again go away upon augmentation of the basis,
requiring no extra considerations for establishing convergence. Also
the neutral atoms show improved accuracy over the UGBS-style UHGBS
basis sets, as the truncation error for superheavy atoms is reduced
roughly by an order of magnitude from tens of m$E_{h}$ for UHGBS-9
to m$E_{h}$ for HGBS-9.

\subsubsection{Molecular DFT benchmarks\label{subsec:Molecular-benchmarks}}

While accuracy on atoms is important, there is often little reason
to use Gaussian basis sets for atomic calculations as numerically
exact calculations can be routinely performed.\citep{Lehtola2019a,Lehtola2019c,Lehtola2019e}
Instead, the acid test for basis sets is their performance on molecules.

A fully numerical benchmark study\citep{Jensen2017} has recently
reported total and atomization energies for 211 molecules with $\mu E_{h}$
accuracy. The study showed that commonly-used Gaussian basis sets
failed to reach chemical accuracy (conventionally defined as an error
smaller than 1 kcal/mol) for several molecules. Followup studies showed
that the correct reproduction of the atomization energies requires
polarization consistent basis sets of at least quadruple-$\zeta$
quality,\citep{Jensen2017b} and correlation consistent basis sets
of at least quintuple-$\zeta$ quality.\citep{Feller2018} To assess
the accuracy of the (a)HGBS basis sets and their polarized counterparts,
we examine a subset of the 211 molecules of \citeref{Jensen2017}:
\ce{N2}, CO, \ce{F2}, \ce{C2H4}, LiF, HF, \ce{H2O}, \ce{NH3},
\ce{CH4}, \ce{P2}, \ce{SiS}, \ce{Cl2}, \ce{Na2}, \ce{SO2}, \ce{AlF3},
\ce{PF5}, and \ce{SF6}. This set contains standard test molecules
for basis set studies, their heavier homologues, as well as the worst
offenders of \citeref{Jensen2017}: the atomization energy of \ce{SF6}
is the worst-case scenario for aug-cc-pV5Z (an error of 2.320 kcal/mol)
and def2-QZVPD (an error of 2.417 kcal/mol), whereas \ce{SO2} is
the worst offender for pc-3 (an error of 0.580 kcal/mol).\citep{Jensen2017}

The molecular calculations are performed with \PsiFour{}\citep{Parrish2017}
and the Perdew--Burke--Ernzerhof (PBE) functional\citep{Perdew1996,Perdew1997}
using a (100,590) quadrature grid and geometries from \citeref{Jensen2017}.
Although linear dependence issues may arise for large molecules in
large basis sets, they can nowadays be routinely resolved via the
recently proposed partial Cholesky orthogonalization procedure.\citep{Lehtola2019f,Lehtola2020a}
A sub-basis set for the molecular calculation is selected from the
full pool of basis functions via a partial Cholesky decomposition
of the overlap matrix to a tight $10^{-9}$ threshold, and the sub-basis
is orthogonalized with the default $10^{-7}$ threshold used in \PsiFour{}.
For instance, in the largest calculation of the present work, \ce{SF6}
in the aHGBSP3-9 basis set, the Cholesky procedure removed 593 linearly
dependent basis functions out of the total 4033 basis functions in
the calculation, after which a further 201 linear combinations of
the remaining basis functions were removed by the canonical orthogonalization
procedure; altogether 19.7\% of the basis functions were thus removed.

\begin{figure*}
\begin{centering}
\subfloat[Normal basis]{\begin{centering}
\includegraphics[width=0.5\textwidth]{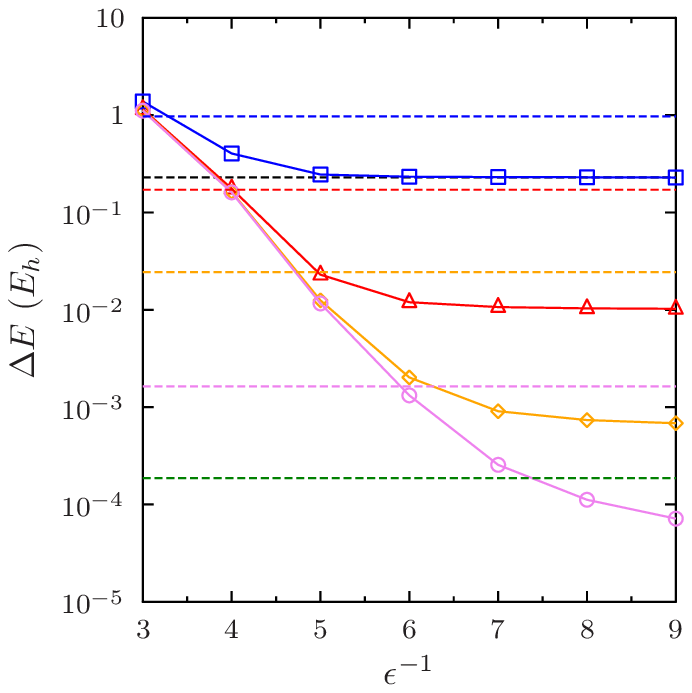}
\par\end{centering}
}\subfloat[Augmented basis]{\begin{centering}
\includegraphics[width=0.5\textwidth]{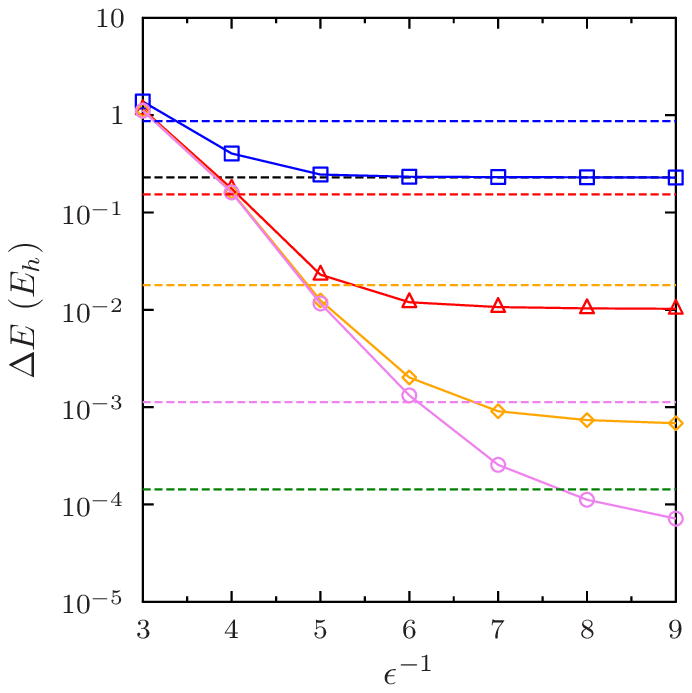}
\par\end{centering}
}
\par\end{centering}
\caption{Truncation error in the total energy of \ce{SO2} with energy-optimized
sets for various thresholds $\epsilon$. See main text for legend.\label{fig:SO2}}
\end{figure*}

\begin{figure*}
\begin{centering}
\subfloat[Normal basis]{\begin{centering}
\includegraphics[width=0.5\textwidth]{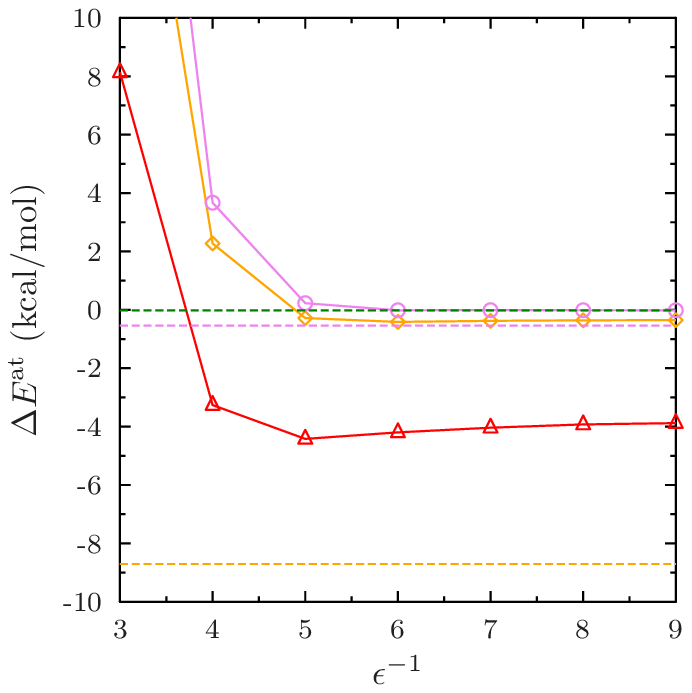}
\par\end{centering}
}\subfloat[Augmented basis]{\begin{centering}
\includegraphics[width=0.5\textwidth]{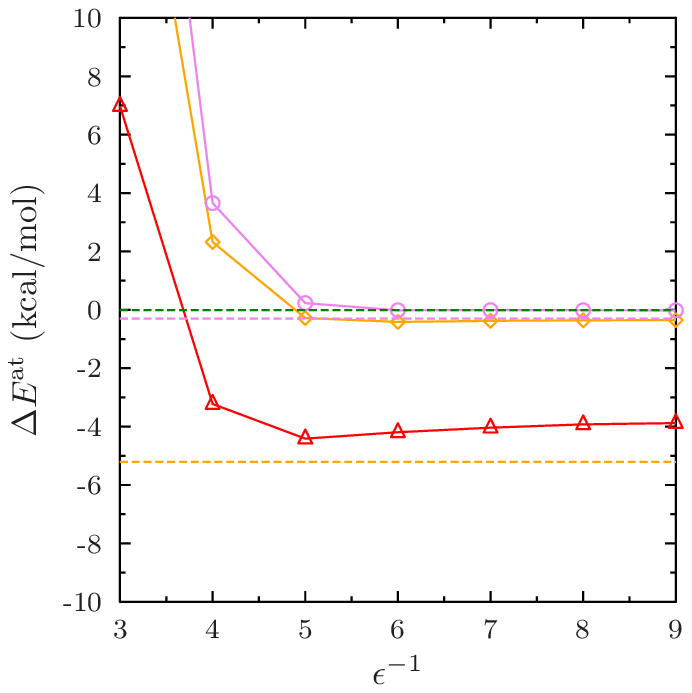}
\par\end{centering}
}
\par\end{centering}
\caption{Truncation error in the atomization energy of \ce{SO2} with energy-optimized
sets for various thresholds $\epsilon$. See main text for legend.
The truncation error for the unpolarized $n=0$ basis sets do not
fit in the graph. \label{fig:SO2-atE}}
\end{figure*}

The results for the total and atomization energy of \ce{SO2} are
shown in \figref{SO2, SO2-atE}, respectively; the rest of the data
behave similarly and can be found in the supporting information. The
plots contain data for the (a)HGBSP$n$-$m$ basis sets for $n=0,1,2,3$
and $m=3,\dots,9$. The $m$ value, which corresponds to the energy
threshold used to form the basis set, is shown on the $x$ axis, whereas
separate curves are shown for the various $n$ values. The polarization-free
$n=0$ data are shown with solid blue lines and blue squares, the
$n=1$ data are shown with solid red lines with red triangles, the
$n=2$ data with orange solid lines and orange diamonds, and the $n=3$
data with violet solid lines and violet circles.

For comparison, the figures also show data for the UGBS basis (horizontal
black dashed line), as well as the un-(aug-)pc-$n$ basis sets. In
analogy to the (a)HGBS data, un-(aug-)pc-$0$ data are shown as the
dashed blue horizontal line, the un-(aug-)pc-$1$ data as the dashed
red horizontal line, the un-(aug-)pc-$2$ data as the dashed orange
line, the un-(aug-)pc-$3$ data as the dashed violet line, and the
un-(aug-)pc-$4$ data as the dashed green horizontal line. The HGBSP0-$m$,
\emph{i.e.}, HGBS-$m$ data tend to agree with UGBS for large $m$
in cases where UGBS is accurate; in other cases the HGBS-$m$ data
undercuts UGBS. In contrast to UGBS, the un-(aug-)pc-$n$ sets are
truncated for optimal balance between errors in atomic and polarization
energies, due to which correspondences between HGBS and un-(aug-)pc-$n$
data cannot be straightforwardly defined.

The results emphatically show the well-known importance of polarization
functions in molecular calculations. For instance, the atomization
energy of \ce{SF6} has an basis set truncation error error of $-275$
kcal/mol in the UGBS basis, and $-266$ kcal/mol in the HGBS-9 basis,
obviously making molecular calculations in such basis sets of little
avail. The truncation error is drastically reduced by the addition
of polarization shells, reducing to $-10.47$ kcal/mol in HGBSP1-9,
to $-0.998$ kcal/mol in HGBSP2-9 and to $-0.036$ kcal/mol in HGBSP3-9.
(For comparison, the truncation error for the atomization energy of
\ce{SF6} is $0.053$ kcal/mol in un-pc-4 and $0.038$ kcal/mol in
un-aug-pc-4.)

\subsubsection{Transition metal diatomics\label{subsec:Transition-metal-diatomics}}

\begin{figure*}
\begin{centering}
\subfloat[Normal basis]{\begin{centering}
\includegraphics[width=0.5\textwidth]{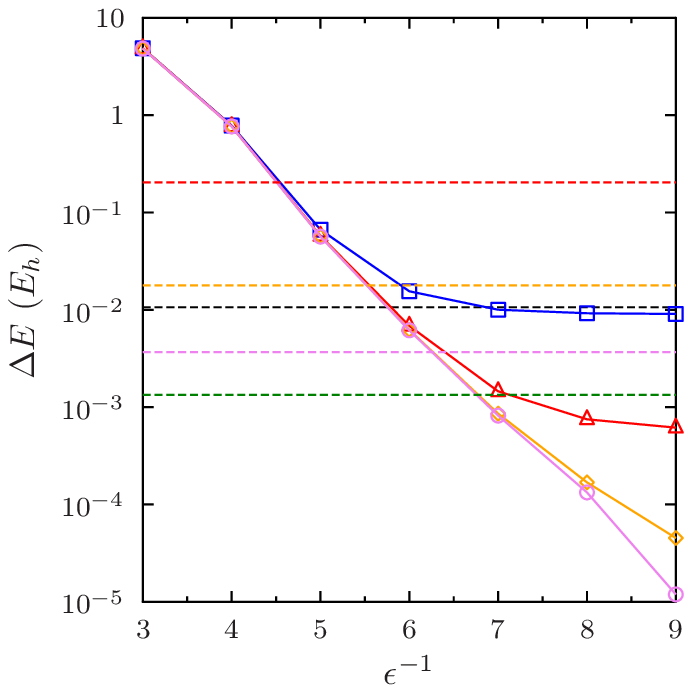}
\par\end{centering}
}\subfloat[Augmented basis]{\begin{centering}
\includegraphics[width=0.5\textwidth]{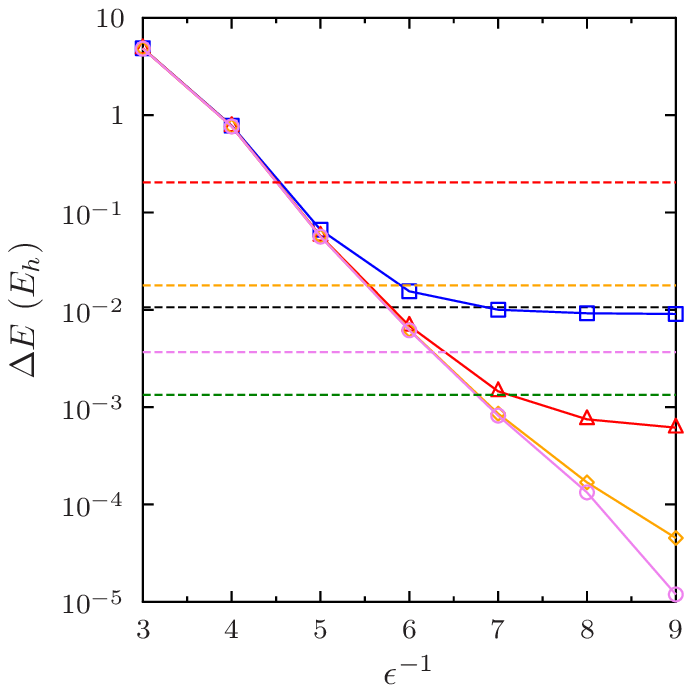}
\par\end{centering}
}
\par\end{centering}
\caption{Truncation error in the total energy of \ce{ZnF} with energy-optimized
sets for various thresholds $\epsilon$. See main text for legend.\label{fig:ZnF}}
\end{figure*}

The density functional calculations on main-group elements are supplemented
with Hartree--Fock calculations on the transition metal diatomics
ScN ($^{1}\Sigma$, $R=1.687$ Å), NiC ($^{1}\Sigma$, $R=1.631$
Å), CuCl ($^{1}\Sigma$, $R=2.051$ Å), and ZnF ($^{2}\Sigma$, $R=1.768$
Å), for which fully numerical reference values have been reported
\citeref{Lehtola2019b}; the calculation for ZnF was spin-unrestricted.
Combining DFT data with Hartree--Fock is reasonable, as the basis
set requirements of Hartree--Fock and density functional calculations
are well-known to be similar. As an example, the convergence of the
total energy of ZnF is shown in \figref{ZnF}; the results for ScN,
NiC, and CuCl are available in the supporting information. The legend
in \figref{ZnF} is the same as in the plots of \subsecref{Molecular-benchmarks};
note, however, that the un-(aug-)pc-0 basis sets are not available
for the transition metals. The results demonstrate that reliable total
energies can be recovered also for transition metal complexes with
the presently developed basis sets; already the (a)HGBS1P-9 basis
set yields a lower total energy than the un-(aug-)pc-4 basis set.

\section{Summary and Conclusions\label{sec:Summary-and-Conclusions}}

We have presented a way to form polarized Gaussian basis sets from
first principles using only one-electron ions, and shown that they
yield accurate electronic structures by (i) non-relativistic spin-restricted
Hartree--Fock calculations with spherically averaged densities on
the neutral atoms $1\leq Z\leq118$ and their cations, (ii) non-relativistic
density-functional total and atomization energies on a set of 17 main-group
molecules, as well as (iii) Hartree--Fock calculations on four first-row
transition metal diatomics. As the method for generating the various
angular momentum shells is unaware of what the basis set is used for,
the method produces polarization and correlation shells in the same
fashion as the shells that are occupied at the self-consistent field
level of theory: in either case, the $l(l+1)/r^{2}$ kinetic energy
barrier makes tight exponents less important in increasing $l$. Arguments
from perturbation theory suggest that tighter exponents than those
reproduced by the current method are likely not important for the
energy even at higher levels of theory.

Since screening effects are included only by the \emph{ad hoc} requirement
that the basis set is accurate for the one-electron ions that interpolate
between the known asymptotics of the optimized effective potential---full
nuclear charge at the nucleus and a $-1/r$ potential far away ($-0.5/r$
for the augmented sets)---the generated basis sets should work well
for whatever purpose. The main drawback of the sets is that they likely
contain unnecessarily many diffuse functions (especially at higher
angular momentum) due to the neglect of electronic screening effects.
However, even significant linear dependencies can be routinely handled
with state-of-the-art approaches,\citep{Lehtola2019f,Lehtola2020a}
and we have shown that the sets achieve the main goal of a simple
way to generate basis sets that allow tractable approaches to the
complete basis set limit. 

Minor problems mostly related to the $ns^{1}$ alkali metals and alkaline
metal cations and a select few other problematic cases were discovered
with the hydrogenic basis sets. It was found that the outliers vanished
upon augmentation of the basis set with diffuse functions, suggesting
that the outermost electrons are moving in a potential weaker than
$-1/r$.

Some large discrepancies were discovered in the atomic calculations
with the UGBS basis set,\citep{DeCastro1998} culminating in a $0.192E_{h}$
truncation error for \ce{Th+}. The UGBS has been parametrized for
atomic multiconfigurational Hartree--Fock ground states, and thereby
has no guarantee of accuracy for other atomic states. Moreover, strictly
speaking UGBS has not even been parametrized as a Gaussian basis set:
\citeref{DeCastro1998} employs a Griffin--Wheeler--Hill\citep{Hill1953,Griffin1957}
(GWH) integral formulation of the Hartree--Fock method,\citep{Mohalem1986}
in which the usual basis set expansion of \eqref{basexp} is replaced
by an integral expression
\begin{equation}
\psi_{i}(\boldsymbol{r})=\int C_{i}(\alpha)\chi(\boldsymbol{r};\alpha){\rm d}\alpha\label{eq:intconv}
\end{equation}
and the unknown weight functions $C_{i}(\alpha)$ are finally solved
at discrete grid-points $\alpha_{j}$. However, as the grid points
$\alpha_{j}$ are typically chosen evenly spaced in $\log\alpha$,
this procedure appears to amount simply to the use of an integral-transform
function basis set,\Citep{Somarjai1968} that is, an even-tempered
set of integrals of Gaussian functions instead of even-tempered Gaussians
as the basis functions. Although the universal hydrogenic basis sets
assessed in the present work employed the same even-tempered expansion
as UGBS, the hydrogenic sets reproduced significantly smaller truncation
errors than UGBS, even though the hydrogenic sets were not optimized
for any particular atomic state.

Polarization functions are of the utmost importance for molecular
applications. As UGBS has been employed in a variety of molecular
calculations, we hope that the presently developed basis sets will
allow better accuracy in benchmark calculations for molecules.

Although only non-relativistic theory has been considered in the present
work, large non-relativistic basis sets have been found to be useful
also for relativistic calculations.\citep{Matsuoka1987} However,
basis sets optimized for the total angular momentum quantum number
$j$ are more efficient for heavy elements,\citep{Dyall1996} requiring
relativistic basis set optimizations. The present scheme can be extended
in a straightforward fashion to relativistic calculations with finite-size
nuclei,\citep{Ishikawa1985} and could be used \eg{} in studies across
and beyond the established periodic table.\citep{Pyykko2011}

\section*{Supporting Information}

See the supporting information for the energy-optimized basis sets
in \textsc{NWChem} and \textsc{Gaussian'94} formats, the full set
of molecular data, as well as the plots for \ce{N2}, CO, \ce{F2},
\ce{C2H4}, LiF, HF, \ce{H2O}, \ce{NH3}, \ce{CH4}, \ce{P2}, \ce{SiS},
\ce{Cl2}, \ce{Na2}, \ce{AlF3}, \ce{PF5}, \ce{SF6}, ScN, NiC,
CuCl, and ZnF.

\section*{Acknowledgments}

This work has been supported by the Academy of Finland (Suomen Akatemia)
through project number 311149. Computational resources provided by
CSC -- It Center for Science Ltd (Espoo, Finland) and the Finnish
Grid and Cloud Infrastructure (persistent identifier urn:nbn:fi:research-infras-2016072533)
are gratefully acknowledged.

\section*{Appendix: completeness-optimized $\beta$}

If necessary, the spacing $\beta$ of the exponents can be derived
from a completeness argument. Assuming an infinitely large even-tempered
expansion $\alpha_{i}=\alpha_{0}\beta^{i}$ with $\alpha_{\min}\to0$
and $\beta\to1$, the largest deviation from completeness $\Delta=1-Y(\alpha)$
will be at $\alpha=\alpha_{0}\beta^{i+1/2}$. The value of $\beta$
can then be fixed based on the maximal deviation from completeness
$\Delta$ at the limit of an infinitely large even-tempered basis
set. However, due to the $l$ dependence of the overlap (see \eqref{overlap}),
the values of $\beta$ decrease monotonically in angular momentum,\citep{Lehtola2015}
leading to a tighter spacing of the exponents for increasing $l$.
In contrast, the basis sets developed in the present work employ a
fixed value of $\beta$ for all angular momenta, as exploratory calculations
suggested the gain in an angular momentum dependent $\beta$ would
be small.

\bibliography{citations}

\end{document}

%% file: HF+.tex
\begin{tabular}{llr|llr|llr}
\ce{He+} & $ 1s^{1} $ & -1.701412 & \ce{Nb+} & $ \textrm{[Kr]} 5s^{2} 4d^{2} $ & -3752.805694 & \ce{Hg+} & $ \textrm{[Xe]} 4f^{14} 5d^{10} 6s^{1} $ & -18408.667135\tabularnewline
\ce{Li+} & $ 1s^{2} $ & -7.236415 & \ce{Mo+} & $ \textrm{[Kr]} 4d^{5} $ & -3974.585529 & \ce{Tl+} & $ \textrm{[Xe]} 4f^{14} 5d^{10} 6s^{2} $ & -18961.638602\tabularnewline
\ce{Be+} & $ \textrm{[He]} 2s^{1} $ & -14.181447 & \ce{Tc+} & $ \textrm{[Kr]} 4d^{6} $ & -4203.848372 & \ce{Pb+} & $ \textrm{[Xe]} 4f^{14} 5d^{10} 6s^{2} 6p^{1} $ & -19523.653096\tabularnewline
\ce{B+} & $ \textrm{[He]} 2s^{2} $ & -24.237575 & \ce{Ru+} & $ \textrm{[Kr]} 4d^{7} $ & -4440.674393 & \ce{Bi+} & $ \textrm{[Xe]} 4f^{14} 5d^{10} 6s^{2} 6p^{2} $ & -20095.066797\tabularnewline
\ce{C+} & $ \textrm{[He]} 2s^{2} 2p^{1} $ & -37.059901 & \ce{Rh+} & $ \textrm{[Kr]} 4d^{8} $ & -4685.158915 & \ce{Po+} & $ \textrm{[Xe]} 4f^{14} 5d^{10} 6s^{2} 6p^{3} $ & -20675.932702\tabularnewline
\ce{N+} & $ \textrm{[He]} 2s^{2} 2p^{2} $ & -53.399066 & \ce{Pd+} & $ \textrm{[Kr]} 4d^{9} $ & -4937.396533 & \ce{At+} & $ \textrm{[Xe]} 4f^{14} 5d^{10} 6s^{2} 6p^{4} $ & -21266.301129\tabularnewline
\ce{O+} & $ \textrm{[He]} 2s^{2} 2p^{3} $ & -73.643807 & \ce{Ag+} & $ \textrm{[Kr]} 4d^{10} $ & -5197.481334 & \ce{Rn+} & $ \textrm{[Xe]} 4f^{14} 5d^{10} 6s^{2} 6p^{5} $ & -21866.221152\tabularnewline
\ce{F+} & $ \textrm{[He]} 2s^{2} 2p^{4} $ & -98.181002 & \ce{Cd+} & $ \textrm{[Kr]} 4d^{10} 5s^{1} $ & -5464.802726 & \ce{Fr+} & $ \textrm{[Xe]} 4f^{14} 5d^{10} 6s^{2} 6p^{6} $ & -22475.741154\tabularnewline
\ce{Ne+} & $ \textrm{[He]} 2s^{2} 2p^{5} $ & -127.396791 & \ce{In+} & $ \textrm{[Kr]} 4d^{10} 5s^{2} $ & -5739.978392 & \ce{Ra+} & $ \textrm{[Rn]} 7s^{1} $ & -23094.116059\tabularnewline
\ce{Na+} & $ \textrm{[He]} 2s^{2} 2p^{6} $ & -161.676963 & \ce{Sn+} & $ \textrm{[Kr]} 4d^{10} 5s^{2} 5p^{1} $ & -6022.562145 & \ce{Ac+} & $ \textrm{[Rn]} 7s^{2} $ & -23721.969484\tabularnewline
\ce{Mg+} & $ \textrm{[Ne]} 3s^{1} $ & -199.294692 & \ce{Sb+} & $ \textrm{[Kr]} 4d^{10} 5s^{2} 5p^{2} $ & -6312.938513 & \ce{Th+} & $ \textrm{[Rn]} 7s^{2} 5f^{1} $ & -24359.219396\tabularnewline
\ce{Al+} & $ \textrm{[Ne]} 3s^{2} $ & -241.674670 & \ce{Te+} & $ \textrm{[Kr]} 4d^{10} 5s^{2} 5p^{3} $ & -6611.181174 & \ce{Pa+} & $ \textrm{[Rn]} 7s^{2} 5f^{2} $ & -25006.262530\tabularnewline
\ce{Si+} & $ \textrm{[Ne]} 3s^{2} 3p^{1} $ & -288.434098 & \ce{I+} & $ \textrm{[Kr]} 4d^{10} 5s^{2} 5p^{4} $ & -6917.361044 & \ce{U+} & $ \textrm{[Rn]} 5f^{4} 7s^{1} $ & -25663.196802\tabularnewline
\ce{P+} & $ \textrm{[Ne]} 3s^{2} 3p^{2} $ & -340.062999 & \ce{Xe+} & $ \textrm{[Kr]} 4d^{10} 5s^{2} 5p^{5} $ & -7231.547777 & \ce{Np+} & $ \textrm{[Rn]} 5f^{6} $ & -26330.186970\tabularnewline
\ce{S+} & $ \textrm{[Ne]} 3s^{2} 3p^{3} $ & -396.750428 & \ce{Cs+} & $ \textrm{[Kr]} 4d^{10} 5s^{2} 5p^{6} $ & -7553.810329 & \ce{Pu+} & $ \textrm{[Rn]} 5f^{7} $ & -27007.176176\tabularnewline
\ce{Cl+} & $ \textrm{[Ne]} 3s^{2} 3p^{4} $ & -458.682406 & \ce{Ba+} & $ \textrm{[Xe]} 6s^{1} $ & -7883.345103 & \ce{Am+} & $ \textrm{[Rn]} 5f^{8} $ & -27694.224176\tabularnewline
\ce{Ar+} & $ \textrm{[Ne]} 3s^{2} 3p^{5} $ & -526.043520 & \ce{La+} & $ \textrm{[Xe]} 6s^{2} $ & -8220.831565 & \ce{Cm+} & $ \textrm{[Rn]} 5f^{9} $ & -28391.390363\tabularnewline
\ce{K+} & $ \textrm{[Ne]} 3s^{2} 3p^{6} $ & -599.017579 & \ce{Ce+} & $ \textrm{[Xe]} 6s^{2} 4f^{1} $ & -8566.210754 & \ce{Bk+} & $ \textrm{[Rn]} 5f^{10} $ & -29098.733569\tabularnewline
\ce{Ca+} & $ \textrm{[Ar]} 4s^{1} $ & -676.511217 & \ce{Pr+} & $ \textrm{[Xe]} 6s^{2} 4f^{2} $ & -8919.987528 & \ce{Cf+} & $ \textrm{[Rn]} 5f^{11} $ & -29816.312145\tabularnewline
\ce{Sc+} & $ \textrm{[Ar]} 4s^{2} $ & -759.462097 & \ce{Nd+} & $ \textrm{[Xe]} 4f^{3} 6s^{2} $ & -9282.254442 & \ce{Es+} & $ \textrm{[Rn]} 5f^{12} $ & -30544.184035\tabularnewline
\ce{Ti+} & $ \textrm{[Ar]} 4s^{2} 3d^{1} $ & -847.770753 & \ce{Pm+} & $ \textrm{[Xe]} 6s^{1} 4f^{5} $ & -9653.144487 & \ce{Fm+} & $ \textrm{[Rn]} 5f^{13} $ & -31282.406851\tabularnewline
\ce{V+} & $ \textrm{[Ar]} 4s^{2} 3d^{2} $ & -941.892534 & \ce{Sm+} & $ \textrm{[Xe]} 4f^{7} $ & -10032.758417 & \ce{Md+} & $ \textrm{[Rn]} 5f^{14} $ & -32031.037923\tabularnewline
\ce{Cr+} & $ \textrm{[Ar]} 3d^{5} $ & -1042.129030 & \ce{Eu+} & $ \textrm{[Xe]} 4f^{8} $ & -10421.155009 & \ce{No+} & $ \textrm{[Rn]} 5f^{14} 7s^{1} $ & -32789.296998\tabularnewline
\ce{Mn+} & $ \textrm{[Ar]} 3d^{6} $ & -1148.539203 & \ce{Gd+} & $ \textrm{[Xe]} 4f^{9} $ & -10818.384689 & \ce{Lr+} & $ \textrm{[Rn]} 5f^{14} 7s^{2} $ & -33557.718309\tabularnewline
\ce{Fe+} & $ \textrm{[Ar]} 3d^{7} $ & -1261.250878 & \ce{Tb+} & $ \textrm{[Xe]} 4f^{10} $ & -11224.528879 & \ce{Rf+} & $ \textrm{[Rn]} 5f^{14} 7s^{2} 6d^{1} $ & -34336.150838\tabularnewline
\ce{Co+} & $ \textrm{[Ar]} 3d^{8} $ & -1380.417722 & \ce{Dy+} & $ \textrm{[Xe]} 4f^{11} $ & -11639.668824 & \ce{Db+} & $ \textrm{[Rn]} 5f^{14} 6d^{4} $ & -35124.903231\tabularnewline
\ce{Ni+} & $ \textrm{[Ar]} 3d^{9} $ & -1506.192686 & \ce{Ho+} & $ \textrm{[Xe]} 4f^{12} $ & -12063.885572 & \ce{Sg+} & $ \textrm{[Rn]} 5f^{14} 6d^{5} $ & -35924.051625\tabularnewline
\ce{Cu+} & $ \textrm{[Ar]} 3d^{10} $ & -1638.728242 & \ce{Er+} & $ \textrm{[Xe]} 4f^{13} $ & -12497.259968 & \ce{Bh+} & $ \textrm{[Rn]} 5f^{14} 6d^{6} $ & -36733.566255\tabularnewline
\ce{Zn+} & $ \textrm{[Ar]} 3d^{10} 4s^{1} $ & -1777.481935 & \ce{Tm+} & $ \textrm{[Xe]} 4f^{14} $ & -12939.872673 & \ce{Hs+} & $ \textrm{[Rn]} 5f^{14} 6d^{7} $ & -37553.490767\tabularnewline
\ce{Ga+} & $ \textrm{[Ar]} 3d^{10} 4s^{2} $ & -1923.059722 & \ce{Yb+} & $ \textrm{[Xe]} 4f^{14} 6s^{1} $ & -13391.225921 & \ce{Mt+} & $ \textrm{[Rn]} 5f^{14} 6d^{8} $ & -38383.868493\tabularnewline
\ce{Ge+} & $ \textrm{[Ar]} 3d^{10} 4s^{2} 4p^{1} $ & -2074.953600 & \ce{Lu+} & $ \textrm{[Xe]} 4f^{14} 6s^{2} $ & -13851.600989 & \ce{Ds+} & $ \textrm{[Rn]} 5f^{14} 6d^{9} $ & -39224.742591\tabularnewline
\ce{As+} & $ \textrm{[Ar]} 3d^{10} 4s^{2} 4p^{2} $ & -2233.624153 & \ce{Hf+} & $ \textrm{[Xe]} 4f^{14} 6s^{2} 5d^{1} $ & -14320.787001 & \ce{Rg+} & $ \textrm{[Rn]} 5f^{14} 6d^{10} $ & -40076.156119\tabularnewline
\ce{Se+} & $ \textrm{[Ar]} 3d^{10} 4s^{2} 4p^{3} $ & -2399.180057 & \ce{Ta+} & $ \textrm{[Xe]} 4f^{14} 6s^{2} 5d^{2} $ & -14799.101433 & \ce{Cn+} & $ \textrm{[Rn]} 5f^{14} 6d^{10} 7s^{1} $ & -40937.501696\tabularnewline
\ce{Br+} & $ \textrm{[Ar]} 3d^{10} 4s^{2} 4p^{4} $ & -2571.727705 & \ce{W+} & $ \textrm{[Xe]} 4f^{14} 5d^{5} $ & -15286.728283 & \ce{Nh+} & $ \textrm{[Rn]} 5f^{14} 6d^{10} 7s^{2} $ & -41809.357021\tabularnewline
\ce{Kr+} & $ \textrm{[Ar]} 3d^{10} 4s^{2} 4p^{5} $ & -2751.372626 & \ce{Re+} & $ \textrm{[Xe]} 4f^{14} 5d^{6} $ & -15783.648951 & \ce{Fl+} & $ \textrm{[Rn]} 5f^{14} 6d^{10} 7s^{2} 7p^{1} $ & -42691.324053\tabularnewline
\ce{Rb+} & $ \textrm{[Ar]} 3d^{10} 4s^{2} 4p^{6} $ & -2938.219931 & \ce{Os+} & $ \textrm{[Xe]} 4f^{14} 5d^{7} $ & -16289.894965 & \ce{Mc+} & $ \textrm{[Rn]} 5f^{14} 6d^{10} 7s^{2} 7p^{2} $ & -43583.716355\tabularnewline
\ce{Sr+} & $ \textrm{[Kr]} 5s^{1} $ & -3131.320472 & \ce{Ir+} & $ \textrm{[Xe]} 4f^{14} 5d^{8} $ & -16805.526014 & \ce{Lv+} & $ \textrm{[Rn]} 5f^{14} 6d^{10} 7s^{2} 7p^{3} $ & -44486.575688\tabularnewline
\ce{Y+} & $ \textrm{[Kr]} 5s^{2} $ & -3331.472882 & \ce{Pt+} & $ \textrm{[Xe]} 4f^{14} 5d^{9} $ & -17330.601930 & \ce{Ts+} & $ \textrm{[Rn]} 5f^{14} 6d^{10} 7s^{2} 7p^{4} $ & -45399.941031\tabularnewline
\ce{Zr+} & $ \textrm{[Kr]} 5s^{2} 4d^{1} $ & -3538.516109 & \ce{Au+} & $ \textrm{[Xe]} 4f^{14} 5d^{10} $ & -17865.182681 & \ce{Og+} & $ \textrm{[Rn]} 5f^{14} 6d^{10} 7s^{2} 7p^{5} $ & -46323.849966\tabularnewline
\end{tabular}

%% file: complete_E.bbl
\begin{thebibliography}{75}%
\makeatletter
\providecommand \@ifxundefined [1]{%
 \@ifx{#1\undefined}
}%
\providecommand \@ifnum [1]{%
 \ifnum #1\expandafter \@firstoftwo
 \else \expandafter \@secondoftwo
 \fi
}%
\providecommand \@ifx [1]{%
 \ifx #1\expandafter \@firstoftwo
 \else \expandafter \@secondoftwo
 \fi
}%
\providecommand \natexlab [1]{#1}%
\providecommand \enquote  [1]{``#1''}%
\providecommand \bibnamefont  [1]{#1}%
\providecommand \bibfnamefont [1]{#1}%
\providecommand \citenamefont [1]{#1}%
\providecommand \href@noop [0]{\@secondoftwo}%
\providecommand \href [0]{\begingroup \@sanitize@url \@href}%
\providecommand \@href[1]{\@@startlink{#1}\@@href}%
\providecommand \@@href[1]{\endgroup#1\@@endlink}%
\providecommand \@sanitize@url [0]{\catcode `\\12\catcode `\$12\catcode
  `\&12\catcode `\#12\catcode `\^12\catcode `\_12\catcode `\%12\relax}%
\providecommand \@@startlink[1]{}%
\providecommand \@@endlink[0]{}%
\providecommand \url  [0]{\begingroup\@sanitize@url \@url }%
\providecommand \@url [1]{\endgroup\@href {#1}{\urlprefix }}%
\providecommand \urlprefix  [0]{URL }%
\providecommand \Eprint [0]{\href }%
\providecommand \doibase [0]{http://dx.doi.org/}%
\providecommand \selectlanguage [0]{\@gobble}%
\providecommand \bibinfo  [0]{\@secondoftwo}%
\providecommand \bibfield  [0]{\@secondoftwo}%
\providecommand \translation [1]{[#1]}%
\providecommand \BibitemOpen [0]{}%
\providecommand \bibitemStop [0]{}%
\providecommand \bibitemNoStop [0]{.\EOS\space}%
\providecommand \EOS [0]{\spacefactor3000\relax}%
\providecommand \BibitemShut  [1]{\csname bibitem#1\endcsname}%
\let\auto@bib@innerbib\@empty
\bibitem [{\citenamefont {Lehtola}(2019{\natexlab{a}})}]{Lehtola2019c}%
  \BibitemOpen
  \bibfield  {author} {\bibinfo {author} {\bibfnamefont {S.}~\bibnamefont
  {Lehtola}},\ }\bibfield  {title} {\enquote {\bibinfo {title} {{A review on
  non-relativistic, fully numerical electronic structure calculations on atoms
  and diatomic molecules}},}\ }\href {\doibase 10.1002/qua.25968} {\bibfield
  {journal} {\bibinfo  {journal} {Int. J. Quantum Chem.}\ }\textbf {\bibinfo
  {volume} {119}},\ \bibinfo {pages} {e25968} (\bibinfo {year}
  {2019}{\natexlab{a}})},\ \Eprint {http://arxiv.org/abs/1902.01431}
  {arXiv:1902.01431} \BibitemShut {NoStop}%
\bibitem [{\citenamefont {Davidson}\ and\ \citenamefont
  {Feller}(1986)}]{Davidson1986}%
  \BibitemOpen
  \bibfield  {author} {\bibinfo {author} {\bibfnamefont {E.~R.}\ \bibnamefont
  {Davidson}}\ and\ \bibinfo {author} {\bibfnamefont {D.}~\bibnamefont
  {Feller}},\ }\bibfield  {title} {\enquote {\bibinfo {title} {{Basis set
  selection for molecular calculations}},}\ }\href {\doibase
  10.1021/cr00074a002} {\bibfield  {journal} {\bibinfo  {journal} {Chem. Rev.}\
  }\textbf {\bibinfo {volume} {86}},\ \bibinfo {pages} {681--696} (\bibinfo
  {year} {1986})}\BibitemShut {NoStop}%
\bibitem [{\citenamefont {Jensen}(2013)}]{Jensen2013b}%
  \BibitemOpen
  \bibfield  {author} {\bibinfo {author} {\bibfnamefont {F.}~\bibnamefont
  {Jensen}},\ }\bibfield  {title} {\enquote {\bibinfo {title} {{Atomic orbital
  basis sets}},}\ }\href {\doibase 10.1002/wcms.1123} {\bibfield  {journal}
  {\bibinfo  {journal} {Wiley Interdiscip. Rev. Comput. Mol. Sci.}\ }\textbf
  {\bibinfo {volume} {3}},\ \bibinfo {pages} {273--295} (\bibinfo {year}
  {2013})}\BibitemShut {NoStop}%
\bibitem [{\citenamefont {Hill}(2013)}]{Hill2013}%
  \BibitemOpen
  \bibfield  {author} {\bibinfo {author} {\bibfnamefont {J.~G.}\ \bibnamefont
  {Hill}},\ }\bibfield  {title} {\enquote {\bibinfo {title} {{Gaussian basis
  sets for molecular applications}},}\ }\href {\doibase 10.1002/qua.24355}
  {\bibfield  {journal} {\bibinfo  {journal} {Int. J. Quantum Chem.}\ }\textbf
  {\bibinfo {volume} {113}},\ \bibinfo {pages} {21--34} (\bibinfo {year}
  {2013})}\BibitemShut {NoStop}%
\bibitem [{\citenamefont {Roos}\ \emph {et~al.}(2004)\citenamefont {Roos},
  \citenamefont {Lindh}, \citenamefont {Malmqvist}, \citenamefont {Veryazov},\
  and\ \citenamefont {Widmark}}]{Roos2004}%
  \BibitemOpen
  \bibfield  {author} {\bibinfo {author} {\bibfnamefont {B.~O.}\ \bibnamefont
  {Roos}}, \bibinfo {author} {\bibfnamefont {R.}~\bibnamefont {Lindh}},
  \bibinfo {author} {\bibfnamefont {P.-{\AA}.}\ \bibnamefont {Malmqvist}},
  \bibinfo {author} {\bibfnamefont {V.}~\bibnamefont {Veryazov}}, \ and\
  \bibinfo {author} {\bibfnamefont {P.-O.}\ \bibnamefont {Widmark}},\
  }\bibfield  {title} {\enquote {\bibinfo {title} {{Main Group Atoms and Dimers
  Studied with a New Relativistic ANO Basis Set}},}\ }\href {\doibase
  10.1021/jp031064+} {\bibfield  {journal} {\bibinfo  {journal} {J. Phys. Chem.
  A}\ }\textbf {\bibinfo {volume} {108}},\ \bibinfo {pages} {2851--2858}
  (\bibinfo {year} {2004})}\BibitemShut {NoStop}%
\bibitem [{\citenamefont {Pollak}\ and\ \citenamefont
  {Weigend}(2017)}]{Pollak2017}%
  \BibitemOpen
  \bibfield  {author} {\bibinfo {author} {\bibfnamefont {P.}~\bibnamefont
  {Pollak}}\ and\ \bibinfo {author} {\bibfnamefont {F.}~\bibnamefont
  {Weigend}},\ }\bibfield  {title} {\enquote {\bibinfo {title} {{Segmented
  Contracted Error-Consistent Basis Sets of Double- and Triple-$\zeta$ Valence
  Quality for One- and Two-Component Relativistic All-Electron
  Calculations}},}\ }\href {\doibase 10.1021/acs.jctc.7b00593} {\bibfield
  {journal} {\bibinfo  {journal} {J. Chem. Theory Comput.}\ }\textbf {\bibinfo
  {volume} {13}},\ \bibinfo {pages} {3696--3705} (\bibinfo {year}
  {2017})}\BibitemShut {NoStop}%
\bibitem [{\citenamefont {Zobel}, \citenamefont {Widmark},\ and\ \citenamefont
  {Veryazov}(2020)}]{Zobel2020}%
  \BibitemOpen
  \bibfield  {author} {\bibinfo {author} {\bibfnamefont {J.~P.}\ \bibnamefont
  {Zobel}}, \bibinfo {author} {\bibfnamefont {P.-O.}\ \bibnamefont {Widmark}},
  \ and\ \bibinfo {author} {\bibfnamefont {V.}~\bibnamefont {Veryazov}},\
  }\bibfield  {title} {\enquote {\bibinfo {title} {{The ANO-R Basis Set}},}\
  }\href {\doibase 10.1021/acs.jctc.9b00873} {\bibfield  {journal} {\bibinfo
  {journal} {J. Chem. Theory Comput.}\ }\textbf {\bibinfo {volume} {16}},\
  \bibinfo {pages} {278--294} (\bibinfo {year} {2020})}\BibitemShut {NoStop}%
\bibitem [{\citenamefont {Silver}\ and\ \citenamefont
  {Nieuwpoort}(1978)}]{Silver1978}%
  \BibitemOpen
  \bibfield  {author} {\bibinfo {author} {\bibfnamefont {D.~M.}\ \bibnamefont
  {Silver}}\ and\ \bibinfo {author} {\bibfnamefont {W.~C.}\ \bibnamefont
  {Nieuwpoort}},\ }\bibfield  {title} {\enquote {\bibinfo {title} {{Universal
  atomic basis sets}},}\ }\href {\doibase 10.1016/0009-2614(78)85539-0}
  {\bibfield  {journal} {\bibinfo  {journal} {Chem. Phys. Lett.}\ }\textbf
  {\bibinfo {volume} {57}},\ \bibinfo {pages} {421--422} (\bibinfo {year}
  {1978})}\BibitemShut {NoStop}%
\bibitem [{\citenamefont {Silver}, \citenamefont {Wilson},\ and\ \citenamefont
  {Nieuwpoort}(1978)}]{Silver1978a}%
  \BibitemOpen
  \bibfield  {author} {\bibinfo {author} {\bibfnamefont {D.~M.}\ \bibnamefont
  {Silver}}, \bibinfo {author} {\bibfnamefont {S.}~\bibnamefont {Wilson}}, \
  and\ \bibinfo {author} {\bibfnamefont {W.~C.}\ \bibnamefont {Nieuwpoort}},\
  }\bibfield  {title} {\enquote {\bibinfo {title} {{Universal basis sets and
  transferability of integrals}},}\ }\href {\doibase 10.1002/qua.560140510}
  {\bibfield  {journal} {\bibinfo  {journal} {Int. J. Quantum Chem.}\ }\textbf
  {\bibinfo {volume} {14}},\ \bibinfo {pages} {635--639} (\bibinfo {year}
  {1978})}\BibitemShut {NoStop}%
\bibitem [{\citenamefont {Lehtola}(2019{\natexlab{b}})}]{Lehtola2019a}%
  \BibitemOpen
  \bibfield  {author} {\bibinfo {author} {\bibfnamefont {S.}~\bibnamefont
  {Lehtola}},\ }\bibfield  {title} {\enquote {\bibinfo {title} {{Fully
  numerical Hartree--Fock and density functional calculations. I. Atoms}},}\
  }\href {\doibase 10.1002/qua.25945} {\bibfield  {journal} {\bibinfo
  {journal} {Int. J. Quantum Chem.}\ }\textbf {\bibinfo {volume} {119}},\
  \bibinfo {pages} {e25945} (\bibinfo {year} {2019}{\natexlab{b}})},\ \Eprint
  {http://arxiv.org/abs/1810.11651} {arXiv:1810.11651} \BibitemShut {NoStop}%
\bibitem [{\citenamefont {Lehtola}(2019{\natexlab{c}})}]{Lehtola2019b}%
  \BibitemOpen
  \bibfield  {author} {\bibinfo {author} {\bibfnamefont {S.}~\bibnamefont
  {Lehtola}},\ }\bibfield  {title} {\enquote {\bibinfo {title} {{Fully
  numerical Hartree--Fock and density functional calculations. II. Diatomic
  molecules}},}\ }\href {\doibase 10.1002/qua.25944} {\bibfield  {journal}
  {\bibinfo  {journal} {Int. J. Quantum Chem.}\ }\textbf {\bibinfo {volume}
  {119}},\ \bibinfo {pages} {e25944} (\bibinfo {year} {2019}{\natexlab{c}})},\
  \Eprint {http://arxiv.org/abs/1810.11653} {arXiv:1810.11653} \BibitemShut
  {NoStop}%
\bibitem [{\citenamefont {Lehtola}(2020)}]{Lehtola2019e}%
  \BibitemOpen
  \bibfield  {author} {\bibinfo {author} {\bibfnamefont {S.}~\bibnamefont
  {Lehtola}},\ }\bibfield  {title} {\enquote {\bibinfo {title} {{Fully
  numerical calculations on atoms with fractional occupations and
  range-separated exchange functionals}},}\ }\href {\doibase
  10.1103/PhysRevA.101.012516} {\bibfield  {journal} {\bibinfo  {journal}
  {Phys. Rev. A}\ }\textbf {\bibinfo {volume} {101}},\ \bibinfo {pages}
  {012516} (\bibinfo {year} {2020})},\ \Eprint
  {http://arxiv.org/abs/1908.02528} {arXiv:1908.02528} \BibitemShut {NoStop}%
\bibitem [{\citenamefont {de~Castro}\ and\ \citenamefont
  {Jorge}(1998)}]{DeCastro1998}%
  \BibitemOpen
  \bibfield  {author} {\bibinfo {author} {\bibfnamefont {E.~V.~R.}\
  \bibnamefont {de~Castro}}\ and\ \bibinfo {author} {\bibfnamefont {F.~E.}\
  \bibnamefont {Jorge}},\ }\bibfield  {title} {\enquote {\bibinfo {title}
  {{Accurate universal Gaussian basis set for all atoms of the Periodic
  Table}},}\ }\href {\doibase 10.1063/1.475959} {\bibfield  {journal} {\bibinfo
   {journal} {J. Chem. Phys.}\ }\textbf {\bibinfo {volume} {108}},\ \bibinfo
  {pages} {5225} (\bibinfo {year} {1998})}\BibitemShut {NoStop}%
\bibitem [{\citenamefont {Mohallem}\ and\ \citenamefont
  {Trsic}(1987)}]{Mohallem1987}%
  \BibitemOpen
  \bibfield  {author} {\bibinfo {author} {\bibfnamefont {J.~R.}\ \bibnamefont
  {Mohallem}}\ and\ \bibinfo {author} {\bibfnamefont {M.}~\bibnamefont
  {Trsic}},\ }\bibfield  {title} {\enquote {\bibinfo {title} {{A universal
  Gaussian basis set for atoms Li through Ne based on a generator coordinate
  version of the Hartree--Fock equations}},}\ }\href {\doibase
  10.1063/1.452680} {\bibfield  {journal} {\bibinfo  {journal} {J. Chem.
  Phys.}\ }\textbf {\bibinfo {volume} {86}},\ \bibinfo {pages} {5043} (\bibinfo
  {year} {1987})}\BibitemShut {NoStop}%
\bibitem [{\citenamefont {da~Costa}, \citenamefont {Trsic},\ and\ \citenamefont
  {Mohallem}(1987)}]{DaCosta1987}%
  \BibitemOpen
  \bibfield  {author} {\bibinfo {author} {\bibfnamefont {H.~F.~M.}\
  \bibnamefont {da~Costa}}, \bibinfo {author} {\bibfnamefont {M.}~\bibnamefont
  {Trsic}}, \ and\ \bibinfo {author} {\bibfnamefont {J.~R.}\ \bibnamefont
  {Mohallem}},\ }\bibfield  {title} {\enquote {\bibinfo {title} {{Universal
  gaussian and Slater-type basis sets for atoms He to Ar based on an integral
  version of the Hartree--Fock equations}},}\ }\href {\doibase
  10.1080/00268978700102071} {\bibfield  {journal} {\bibinfo  {journal} {Mol.
  Phys.}\ }\textbf {\bibinfo {volume} {62}},\ \bibinfo {pages} {91--95}
  (\bibinfo {year} {1987})}\BibitemShut {NoStop}%
\bibitem [{\citenamefont {da~Silva}, \citenamefont {da~Costa},\ and\
  \citenamefont {Trsic}(1989)}]{DaSilva1989}%
  \BibitemOpen
  \bibfield  {author} {\bibinfo {author} {\bibfnamefont {A.~B.~F.}\
  \bibnamefont {da~Silva}}, \bibinfo {author} {\bibfnamefont {H.~F.~M.}\
  \bibnamefont {da~Costa}}, \ and\ \bibinfo {author} {\bibfnamefont
  {M.}~\bibnamefont {Trsic}},\ }\bibfield  {title} {\enquote {\bibinfo {title}
  {{Universal gaussian and Slater-type bases for atoms H to Xe based on the
  generator coordinate Hartree--Fock method}},}\ }\href {\doibase
  10.1080/00268978900102271} {\bibfield  {journal} {\bibinfo  {journal} {Mol.
  Phys.}\ }\textbf {\bibinfo {volume} {68}},\ \bibinfo {pages} {433--445}
  (\bibinfo {year} {1989})}\BibitemShut {NoStop}%
\bibitem [{\citenamefont {Jorge}, \citenamefont {de~Castro},\ and\
  \citenamefont {da~Silva}(1997)}]{Jorge1997}%
  \BibitemOpen
  \bibfield  {author} {\bibinfo {author} {\bibfnamefont {F.~E.}\ \bibnamefont
  {Jorge}}, \bibinfo {author} {\bibfnamefont {E.~V.~R.}\ \bibnamefont
  {de~Castro}}, \ and\ \bibinfo {author} {\bibfnamefont {A.~B.~F.}\
  \bibnamefont {da~Silva}},\ }\bibfield  {title} {\enquote {\bibinfo {title}
  {{Accurate universal Gaussian basis set for hydrogen through lanthanum
  generated with the generator coordinate Hartree--Fock method}},}\ }\href
  {\doibase 10.1016/S0301-0104(97)00013-X} {\bibfield  {journal} {\bibinfo
  {journal} {Chem. Phys.}\ }\textbf {\bibinfo {volume} {216}},\ \bibinfo
  {pages} {317--321} (\bibinfo {year} {1997})}\BibitemShut {NoStop}%
\bibitem [{\citenamefont {Jorge}, \citenamefont {{De Castro}},\ and\
  \citenamefont {{Da Silva}}(1997)}]{Jorge1997a}%
  \BibitemOpen
  \bibfield  {author} {\bibinfo {author} {\bibfnamefont {F.~E.}\ \bibnamefont
  {Jorge}}, \bibinfo {author} {\bibfnamefont {E.~V.~R.}\ \bibnamefont {{De
  Castro}}}, \ and\ \bibinfo {author} {\bibfnamefont {A.~B.~F.}\ \bibnamefont
  {{Da Silva}}},\ }\bibfield  {title} {\enquote {\bibinfo {title} {{A universal
  Gaussian basis set for atoms cerium through lawrencium generated with the
  generator coordinate Hartree--Fock method}},}\ }\href {\doibase
  10.1002/(SICI)1096-987X(199710)18:13<1565::AID-JCC1>3.0.CO;2-P} {\bibfield
  {journal} {\bibinfo  {journal} {J. Comput. Chem.}\ }\textbf {\bibinfo
  {volume} {18}},\ \bibinfo {pages} {1565--1569} (\bibinfo {year}
  {1997})}\BibitemShut {NoStop}%
\bibitem [{\citenamefont {Pritchard}\ \emph {et~al.}(2019)\citenamefont
  {Pritchard}, \citenamefont {Altarawy}, \citenamefont {Didier}, \citenamefont
  {Gibson},\ and\ \citenamefont {Windus}}]{Pritchard2019}%
  \BibitemOpen
  \bibfield  {author} {\bibinfo {author} {\bibfnamefont {B.~P.}\ \bibnamefont
  {Pritchard}}, \bibinfo {author} {\bibfnamefont {D.}~\bibnamefont {Altarawy}},
  \bibinfo {author} {\bibfnamefont {B.}~\bibnamefont {Didier}}, \bibinfo
  {author} {\bibfnamefont {T.~D.}\ \bibnamefont {Gibson}}, \ and\ \bibinfo
  {author} {\bibfnamefont {T.~L.}\ \bibnamefont {Windus}},\ }\bibfield  {title}
  {\enquote {\bibinfo {title} {{New Basis Set Exchange: An Open, Up-to-Date
  Resource for the Molecular Sciences Community}},}\ }\href {\doibase
  10.1021/acs.jcim.9b00725} {\bibfield  {journal} {\bibinfo  {journal} {J.
  Chem. Inf. Model.}\ }\textbf {\bibinfo {volume} {59}},\ \bibinfo {pages}
  {4814--4820} (\bibinfo {year} {2019})}\BibitemShut {NoStop}%
\bibitem [{\citenamefont {Manz}\ and\ \citenamefont {Sholl}(2010)}]{Manz2010}%
  \BibitemOpen
  \bibfield  {author} {\bibinfo {author} {\bibfnamefont {T.~A.}\ \bibnamefont
  {Manz}}\ and\ \bibinfo {author} {\bibfnamefont {D.~S.}\ \bibnamefont
  {Sholl}},\ }\bibfield  {title} {\enquote {\bibinfo {title} {{Chemically
  Meaningful Atomic Charges That Reproduce the Electrostatic Potential in
  Periodic and Nonperiodic Materials}},}\ }\href {\doibase 10.1021/ct100125x}
  {\bibfield  {journal} {\bibinfo  {journal} {J. Chem. Theory Comput.}\
  }\textbf {\bibinfo {volume} {6}},\ \bibinfo {pages} {2455--2468} (\bibinfo
  {year} {2010})}\BibitemShut {NoStop}%
\bibitem [{\citenamefont {Ryabinkin}, \citenamefont {Kananenka},\ and\
  \citenamefont {Staroverov}(2013)}]{Ryabinkin2013}%
  \BibitemOpen
  \bibfield  {author} {\bibinfo {author} {\bibfnamefont {I.~G.}\ \bibnamefont
  {Ryabinkin}}, \bibinfo {author} {\bibfnamefont {A.~A.}\ \bibnamefont
  {Kananenka}}, \ and\ \bibinfo {author} {\bibfnamefont {V.~N.}\ \bibnamefont
  {Staroverov}},\ }\bibfield  {title} {\enquote {\bibinfo {title} {{Accurate
  and Efficient Approximation to the Optimized Effective Potential for
  Exchange}},}\ }\href {\doibase 10.1103/PhysRevLett.111.013001} {\bibfield
  {journal} {\bibinfo  {journal} {Phys. Rev. Lett.}\ }\textbf {\bibinfo
  {volume} {111}},\ \bibinfo {pages} {013001} (\bibinfo {year} {2013})},\
  \Eprint {http://arxiv.org/abs/1306.3603} {arXiv:1306.3603} \BibitemShut
  {NoStop}%
\bibitem [{\citenamefont {Kollmar}\ and\ \citenamefont
  {Neese}(2014)}]{Kollmar2014}%
  \BibitemOpen
  \bibfield  {author} {\bibinfo {author} {\bibfnamefont {C.}~\bibnamefont
  {Kollmar}}\ and\ \bibinfo {author} {\bibfnamefont {F.}~\bibnamefont
  {Neese}},\ }\bibfield  {title} {\enquote {\bibinfo {title} {{The static
  response function in Kohn--Sham theory: An appropriate basis for its matrix
  representation in case of finite AO basis sets}},}\ }\href {\doibase
  10.1063/1.4896897} {\bibfield  {journal} {\bibinfo  {journal} {J. Chem.
  Phys.}\ }\textbf {\bibinfo {volume} {141}},\ \bibinfo {pages} {134106}
  (\bibinfo {year} {2014})}\BibitemShut {NoStop}%
\bibitem [{\citenamefont {Kohut}, \citenamefont {Ryabinkin},\ and\
  \citenamefont {Staroverov}(2014)}]{Kohut2014}%
  \BibitemOpen
  \bibfield  {author} {\bibinfo {author} {\bibfnamefont {S.~V.}\ \bibnamefont
  {Kohut}}, \bibinfo {author} {\bibfnamefont {I.~G.}\ \bibnamefont
  {Ryabinkin}}, \ and\ \bibinfo {author} {\bibfnamefont {V.~N.}\ \bibnamefont
  {Staroverov}},\ }\bibfield  {title} {\enquote {\bibinfo {title} {{Hierarchy
  of model Kohn--Sham potentials for orbital-dependent functionals: A practical
  alternative to the optimized effective potential method}},}\ }\href {\doibase
  10.1063/1.4871500} {\bibfield  {journal} {\bibinfo  {journal} {J. Chem.
  Phys.}\ }\textbf {\bibinfo {volume} {140}} (\bibinfo {year} {2014}),\
  10.1063/1.4871500}\BibitemShut {NoStop}%
\bibitem [{\citenamefont {Gaiduk}, \citenamefont {Ryabinkin},\ and\
  \citenamefont {Staroverov}(2015)}]{Gaiduk2015}%
  \BibitemOpen
  \bibfield  {author} {\bibinfo {author} {\bibfnamefont {A.~P.}\ \bibnamefont
  {Gaiduk}}, \bibinfo {author} {\bibfnamefont {I.~G.}\ \bibnamefont
  {Ryabinkin}}, \ and\ \bibinfo {author} {\bibfnamefont {V.~N.}\ \bibnamefont
  {Staroverov}},\ }\bibfield  {title} {\enquote {\bibinfo {title} {{Modified
  Slater exchange potential with correct uniform electron gas limit}},}\ }\href
  {\doibase 10.1139/cjc-2014-0250} {\bibfield  {journal} {\bibinfo  {journal}
  {Can. J. Chem.}\ }\textbf {\bibinfo {volume} {93}},\ \bibinfo {pages}
  {91--97} (\bibinfo {year} {2015})}\BibitemShut {NoStop}%
\bibitem [{\citenamefont {Ospadov}\ and\ \citenamefont
  {Staroverov}(2018)}]{Ospadov2018}%
  \BibitemOpen
  \bibfield  {author} {\bibinfo {author} {\bibfnamefont {E.}~\bibnamefont
  {Ospadov}}\ and\ \bibinfo {author} {\bibfnamefont {V.~N.}\ \bibnamefont
  {Staroverov}},\ }\bibfield  {title} {\enquote {\bibinfo {title}
  {{Construction of Fermi Potentials from Electronic Wave Functions}},}\ }\href
  {\doibase 10.1021/acs.jctc.8b00490} {\bibfield  {journal} {\bibinfo
  {journal} {J. Chem. Theory Comput.}\ }\textbf {\bibinfo {volume} {14}},\
  \bibinfo {pages} {4246--4253} (\bibinfo {year} {2018})}\BibitemShut {NoStop}%
\bibitem [{\citenamefont {Kanungo}, \citenamefont {Zimmerman},\ and\
  \citenamefont {Gavini}(2019)}]{Kanungo2019}%
  \BibitemOpen
  \bibfield  {author} {\bibinfo {author} {\bibfnamefont {B.}~\bibnamefont
  {Kanungo}}, \bibinfo {author} {\bibfnamefont {P.~M.}\ \bibnamefont
  {Zimmerman}}, \ and\ \bibinfo {author} {\bibfnamefont {V.}~\bibnamefont
  {Gavini}},\ }\bibfield  {title} {\enquote {\bibinfo {title} {{Exact
  exchange-correlation potentials from ground-state electron densities}},}\
  }\href {\doibase 10.1038/s41467-019-12467-0} {\bibfield  {journal} {\bibinfo
  {journal} {Nat. Commun.}\ }\textbf {\bibinfo {volume} {10}},\ \bibinfo
  {pages} {4497} (\bibinfo {year} {2019})}\BibitemShut {NoStop}%
\bibitem [{\citenamefont {Hodgson}\ \emph {et~al.}(2017)\citenamefont
  {Hodgson}, \citenamefont {Kraisler}, \citenamefont {Schild},\ and\
  \citenamefont {Gross}}]{Hodgson2017}%
  \BibitemOpen
  \bibfield  {author} {\bibinfo {author} {\bibfnamefont {M.~J.~P.}\
  \bibnamefont {Hodgson}}, \bibinfo {author} {\bibfnamefont {E.}~\bibnamefont
  {Kraisler}}, \bibinfo {author} {\bibfnamefont {A.}~\bibnamefont {Schild}}, \
  and\ \bibinfo {author} {\bibfnamefont {E.~K.~U.}\ \bibnamefont {Gross}},\
  }\bibfield  {title} {\enquote {\bibinfo {title} {{How Interatomic Steps in
  the Exact Kohn--Sham Potential Relate to Derivative Discontinuities of the
  Energy}},}\ }\href {\doibase 10.1021/acs.jpclett.7b02615} {\bibfield
  {journal} {\bibinfo  {journal} {J. Phys. Chem. Lett.}\ }\textbf {\bibinfo
  {volume} {8}},\ \bibinfo {pages} {5974--5980} (\bibinfo {year} {2017})},\
  \Eprint {http://arxiv.org/abs/1706.00586} {arXiv:1706.00586} \BibitemShut
  {NoStop}%
\bibitem [{\citenamefont {Ospadov}\ \emph {et~al.}(2018)\citenamefont
  {Ospadov}, \citenamefont {Tao}, \citenamefont {Staroverov},\ and\
  \citenamefont {Perdew}}]{Ospadov2018a}%
  \BibitemOpen
  \bibfield  {author} {\bibinfo {author} {\bibfnamefont {E.}~\bibnamefont
  {Ospadov}}, \bibinfo {author} {\bibfnamefont {J.}~\bibnamefont {Tao}},
  \bibinfo {author} {\bibfnamefont {V.~N.}\ \bibnamefont {Staroverov}}, \ and\
  \bibinfo {author} {\bibfnamefont {J.~P.}\ \bibnamefont {Perdew}},\ }\bibfield
   {title} {\enquote {\bibinfo {title} {{Visualizing atomic sizes and molecular
  shapes with the classical turning surface of the Kohn--Sham potential}},}\
  }\href {\doibase 10.1073/pnas.1814300115} {\bibfield  {journal} {\bibinfo
  {journal} {Proc. Natl. Acad. Sci.}\ }\textbf {\bibinfo {volume} {115}},\
  \bibinfo {pages} {E11578--E11585} (\bibinfo {year} {2018})}\BibitemShut
  {NoStop}%
\bibitem [{\citenamefont {Wang}, \citenamefont {Hong},\ and\ \citenamefont
  {Li}(2000)}]{Wang2000a}%
  \BibitemOpen
  \bibfield  {author} {\bibinfo {author} {\bibfnamefont {F.}~\bibnamefont
  {Wang}}, \bibinfo {author} {\bibfnamefont {G.}~\bibnamefont {Hong}}, \ and\
  \bibinfo {author} {\bibfnamefont {L.}~\bibnamefont {Li}},\ }\bibfield
  {title} {\enquote {\bibinfo {title} {{A simplified scheme for relativistic
  density functional computation in the zeroth-order regular approximation}},}\
  }\href {\doibase 10.1016/S0009-2614(99)01245-2} {\bibfield  {journal}
  {\bibinfo  {journal} {Chem. Phys. Lett.}\ }\textbf {\bibinfo {volume}
  {316}},\ \bibinfo {pages} {318--323} (\bibinfo {year} {2000})}\BibitemShut
  {NoStop}%
\bibitem [{\citenamefont {Deng}, \citenamefont {Cheeseman},\ and\ \citenamefont
  {Frisch}(2006)}]{Deng2006}%
  \BibitemOpen
  \bibfield  {author} {\bibinfo {author} {\bibfnamefont {W.}~\bibnamefont
  {Deng}}, \bibinfo {author} {\bibfnamefont {J.~R.}\ \bibnamefont {Cheeseman}},
  \ and\ \bibinfo {author} {\bibfnamefont {M.~J.}\ \bibnamefont {Frisch}},\
  }\bibfield  {title} {\enquote {\bibinfo {title} {{Calculation of Nuclear
  Spin-Spin Coupling Constants of Molecules with First and Second Row Atoms in
  Study of Basis Set Dependence}},}\ }\href {\doibase 10.1021/ct600110u}
  {\bibfield  {journal} {\bibinfo  {journal} {J. Chem. Theory Comput.}\
  }\textbf {\bibinfo {volume} {2}},\ \bibinfo {pages} {1028--1037} (\bibinfo
  {year} {2006})}\BibitemShut {NoStop}%
\bibitem [{\citenamefont {Haunschild}\ and\ \citenamefont
  {Scuseria}(2010)}]{Haunschild2010a}%
  \BibitemOpen
  \bibfield  {author} {\bibinfo {author} {\bibfnamefont {R.}~\bibnamefont
  {Haunschild}}\ and\ \bibinfo {author} {\bibfnamefont {G.~E.}\ \bibnamefont
  {Scuseria}},\ }\bibfield  {title} {\enquote {\bibinfo {title}
  {{Range-separated local hybrids}},}\ }\href {\doibase 10.1063/1.3451078}
  {\bibfield  {journal} {\bibinfo  {journal} {J. Chem. Phys.}\ }\textbf
  {\bibinfo {volume} {132}},\ \bibinfo {pages} {224106} (\bibinfo {year}
  {2010})}\BibitemShut {NoStop}%
\bibitem [{\citenamefont {Carmona-Esp{\'{i}}ndola}\ \emph
  {et~al.}(2015)\citenamefont {Carmona-Esp{\'{i}}ndola}, \citenamefont
  {G{\'{a}}zquez}, \citenamefont {Vela},\ and\ \citenamefont
  {Trickey}}]{Carmona-Espindola2015}%
  \BibitemOpen
  \bibfield  {author} {\bibinfo {author} {\bibfnamefont {J.}~\bibnamefont
  {Carmona-Esp{\'{i}}ndola}}, \bibinfo {author} {\bibfnamefont {J.~L.}\
  \bibnamefont {G{\'{a}}zquez}}, \bibinfo {author} {\bibfnamefont
  {A.}~\bibnamefont {Vela}}, \ and\ \bibinfo {author} {\bibfnamefont {S.~B.}\
  \bibnamefont {Trickey}},\ }\bibfield  {title} {\enquote {\bibinfo {title}
  {{Generalized gradient approximation exchange energy functional with correct
  asymptotic behavior of the corresponding potential}},}\ }\href {\doibase
  10.1063/1.4906606} {\bibfield  {journal} {\bibinfo  {journal} {J. Chem.
  Phys.}\ }\textbf {\bibinfo {volume} {142}},\ \bibinfo {pages} {054105}
  (\bibinfo {year} {2015})}\BibitemShut {NoStop}%
\bibitem [{\citenamefont {Maier}, \citenamefont {Ikabata},\ and\ \citenamefont
  {Nakai}(2019)}]{Maier2019a}%
  \BibitemOpen
  \bibfield  {author} {\bibinfo {author} {\bibfnamefont {T.~M.}\ \bibnamefont
  {Maier}}, \bibinfo {author} {\bibfnamefont {Y.}~\bibnamefont {Ikabata}}, \
  and\ \bibinfo {author} {\bibfnamefont {H.}~\bibnamefont {Nakai}},\ }\bibfield
   {title} {\enquote {\bibinfo {title} {{Efficient Semi-Numerical
  Implementation of Relativistic Exact Exchange within the Infinite-Order
  Two-Component Method Using a Modified Chain-of-Spheres Method}},}\ }\href
  {\doibase 10.1021/acs.jctc.9b00228} {\bibfield  {journal} {\bibinfo
  {journal} {J. Chem. Theory Comput.}\ }\textbf {\bibinfo {volume} {15}},\
  \bibinfo {pages} {4745--4763} (\bibinfo {year} {2019})}\BibitemShut {NoStop}%
\bibitem [{\citenamefont {Lehtola}(2019{\natexlab{d}})}]{Lehtola2020a}%
  \BibitemOpen
  \bibfield  {author} {\bibinfo {author} {\bibfnamefont {S.}~\bibnamefont
  {Lehtola}},\ }\bibfield  {title} {\enquote {\bibinfo {title} {{On the
  accurate reproduction of strongly repulsive interatomic potentials}},}\
  }\href {http://arxiv.org/abs/1912.12624} {\  (\bibinfo {year}
  {2019}{\natexlab{d}})},\ \Eprint {http://arxiv.org/abs/1912.12624}
  {arXiv:1912.12624} \BibitemShut {NoStop}%
\bibitem [{\citenamefont {Pantazis}\ and\ \citenamefont
  {Neese}(2009)}]{Pantazis2009}%
  \BibitemOpen
  \bibfield  {author} {\bibinfo {author} {\bibfnamefont {D.~A.}\ \bibnamefont
  {Pantazis}}\ and\ \bibinfo {author} {\bibfnamefont {F.}~\bibnamefont
  {Neese}},\ }\bibfield  {title} {\enquote {\bibinfo {title} {{All-Electron
  Scalar Relativistic Basis Sets for the Lanthanides}},}\ }\href {\doibase
  10.1021/ct900090f} {\bibfield  {journal} {\bibinfo  {journal} {J. Chem.
  Theory Comput.}\ }\textbf {\bibinfo {volume} {5}},\ \bibinfo {pages}
  {2229--2238} (\bibinfo {year} {2009})}\BibitemShut {NoStop}%
\bibitem [{\citenamefont {Pantazis}\ and\ \citenamefont
  {Neese}(2011)}]{Pantazis2011}%
  \BibitemOpen
  \bibfield  {author} {\bibinfo {author} {\bibfnamefont {D.~A.}\ \bibnamefont
  {Pantazis}}\ and\ \bibinfo {author} {\bibfnamefont {F.}~\bibnamefont
  {Neese}},\ }\bibfield  {title} {\enquote {\bibinfo {title} {{All-Electron
  Scalar Relativistic Basis Sets for the Actinides}},}\ }\href {\doibase
  10.1021/ct100736b} {\bibfield  {journal} {\bibinfo  {journal} {J. Chem.
  Theory Comput.}\ }\textbf {\bibinfo {volume} {7}},\ \bibinfo {pages}
  {677--684} (\bibinfo {year} {2011})}\BibitemShut {NoStop}%
\bibitem [{\citenamefont {Pantazis}\ and\ \citenamefont
  {Neese}(2012)}]{Pantazis2012}%
  \BibitemOpen
  \bibfield  {author} {\bibinfo {author} {\bibfnamefont {D.~A.}\ \bibnamefont
  {Pantazis}}\ and\ \bibinfo {author} {\bibfnamefont {F.}~\bibnamefont
  {Neese}},\ }\bibfield  {title} {\enquote {\bibinfo {title} {{All-electron
  scalar relativistic basis sets for the 6p elements}},}\ }\href {\doibase
  10.1007/s00214-012-1292-x} {\bibfield  {journal} {\bibinfo  {journal} {Theor.
  Chem. Acc.}\ }\textbf {\bibinfo {volume} {131}},\ \bibinfo {pages} {1292}
  (\bibinfo {year} {2012})}\BibitemShut {NoStop}%
\bibitem [{\citenamefont {Pantazis}\ and\ \citenamefont
  {Neese}(2014)}]{Pantazis2014}%
  \BibitemOpen
  \bibfield  {author} {\bibinfo {author} {\bibfnamefont {D.~A.}\ \bibnamefont
  {Pantazis}}\ and\ \bibinfo {author} {\bibfnamefont {F.}~\bibnamefont
  {Neese}},\ }\bibfield  {title} {\enquote {\bibinfo {title} {{All-electron
  basis sets for heavy elements}},}\ }\href {\doibase 10.1002/wcms.1177}
  {\bibfield  {journal} {\bibinfo  {journal} {Wiley Interdiscip. Rev. Comput.
  Mol. Sci.}\ }\textbf {\bibinfo {volume} {4}},\ \bibinfo {pages} {363--374}
  (\bibinfo {year} {2014})}\BibitemShut {NoStop}%
\bibitem [{\citenamefont {Kaufmann}, \citenamefont {Baumeister},\ and\
  \citenamefont {Jungen}(1989)}]{Kaufmann1989}%
  \BibitemOpen
  \bibfield  {author} {\bibinfo {author} {\bibfnamefont {K.}~\bibnamefont
  {Kaufmann}}, \bibinfo {author} {\bibfnamefont {W.}~\bibnamefont
  {Baumeister}}, \ and\ \bibinfo {author} {\bibfnamefont {M.}~\bibnamefont
  {Jungen}},\ }\bibfield  {title} {\enquote {\bibinfo {title} {{Universal
  Gaussian basis sets for an optimum representation of Rydberg and continuum
  wavefunctions}},}\ }\href {\doibase 10.1088/0953-4075/22/14/007} {\bibfield
  {journal} {\bibinfo  {journal} {J. Phys. B At. Mol. Opt. Phys.}\ }\textbf
  {\bibinfo {volume} {22}},\ \bibinfo {pages} {2223--2240} (\bibinfo {year}
  {1989})}\BibitemShut {NoStop}%
\bibitem [{\citenamefont {Malli}, \citenamefont {{Da Silva}},\ and\
  \citenamefont {Ishikawa}(1993)}]{Malli1993}%
  \BibitemOpen
  \bibfield  {author} {\bibinfo {author} {\bibfnamefont {G.}~\bibnamefont
  {Malli}}, \bibinfo {author} {\bibfnamefont {A.}~\bibnamefont {{Da Silva}}}, \
  and\ \bibinfo {author} {\bibfnamefont {Y.}~\bibnamefont {Ishikawa}},\
  }\bibfield  {title} {\enquote {\bibinfo {title} {{Universal Gaussian basis
  set for accurate ab initio relativistic Dirac--Fock calculations}},}\ }\href
  {\doibase 10.1103/PhysRevA.47.143} {\bibfield  {journal} {\bibinfo  {journal}
  {Phys. Rev. A}\ }\textbf {\bibinfo {volume} {47}},\ \bibinfo {pages}
  {143--146} (\bibinfo {year} {1993})}\BibitemShut {NoStop}%
\bibitem [{\citenamefont {Malli}, \citenamefont {{Da Silva}},\ and\
  \citenamefont {Ishikawa}(1994)}]{Malli1994}%
  \BibitemOpen
  \bibfield  {author} {\bibinfo {author} {\bibfnamefont {G.~L.}\ \bibnamefont
  {Malli}}, \bibinfo {author} {\bibfnamefont {A.~B.~F.}\ \bibnamefont {{Da
  Silva}}}, \ and\ \bibinfo {author} {\bibfnamefont {Y.}~\bibnamefont
  {Ishikawa}},\ }\bibfield  {title} {\enquote {\bibinfo {title} {{Highly
  accurate relativistic universal Gaussian basis set: Dirac--Fock--Coulomb
  calculations for atomic systems up to nobelium}},}\ }\href {\doibase
  10.1063/1.468311} {\bibfield  {journal} {\bibinfo  {journal} {J. Chem.
  Phys.}\ }\textbf {\bibinfo {volume} {101}},\ \bibinfo {pages} {6829}
  (\bibinfo {year} {1994})}\BibitemShut {NoStop}%
\bibitem [{\citenamefont {Haiduke}, \citenamefont {{De Macedo}},\ and\
  \citenamefont {{Da Silva}}(2005)}]{Haiduke2005}%
  \BibitemOpen
  \bibfield  {author} {\bibinfo {author} {\bibfnamefont {R.~L.~A.}\
  \bibnamefont {Haiduke}}, \bibinfo {author} {\bibfnamefont {L.~G.~M.}\
  \bibnamefont {{De Macedo}}}, \ and\ \bibinfo {author} {\bibfnamefont
  {A.~B.~F.}\ \bibnamefont {{Da Silva}}},\ }\bibfield  {title} {\enquote
  {\bibinfo {title} {{An accurate relativistic universal Gaussian basis set for
  hydrogen through Nobelium without variational prolapse and to be used with
  both uniform sphere and Gaussian nucleus models.}}}\ }\href {\doibase
  10.1002/jcc.20223} {\bibfield  {journal} {\bibinfo  {journal} {J. Comput.
  Chem.}\ }\textbf {\bibinfo {volume} {26}},\ \bibinfo {pages} {932--40}
  (\bibinfo {year} {2005})}\BibitemShut {NoStop}%
\bibitem [{\citenamefont {Lehtola}(2019{\natexlab{e}})}]{Lehtola2019f}%
  \BibitemOpen
  \bibfield  {author} {\bibinfo {author} {\bibfnamefont {S.}~\bibnamefont
  {Lehtola}},\ }\bibfield  {title} {\enquote {\bibinfo {title} {{Curing basis
  set overcompleteness with pivoted Cholesky decompositions}},}\ }\href
  {\doibase 10.1063/1.5139948} {\bibfield  {journal} {\bibinfo  {journal} {J.
  Chem. Phys.}\ }\textbf {\bibinfo {volume} {151}},\ \bibinfo {pages} {241102}
  (\bibinfo {year} {2019}{\natexlab{e}})},\ \Eprint
  {http://arxiv.org/abs/1911.10372} {arXiv:1911.10372} \BibitemShut {NoStop}%
\bibitem [{\citenamefont {Hohenberg}\ and\ \citenamefont
  {Kohn}(1964)}]{Hohenberg1964}%
  \BibitemOpen
  \bibfield  {author} {\bibinfo {author} {\bibfnamefont {P.}~\bibnamefont
  {Hohenberg}}\ and\ \bibinfo {author} {\bibfnamefont {W.}~\bibnamefont
  {Kohn}},\ }\bibfield  {title} {\enquote {\bibinfo {title} {{Inhomogeneous
  Electron Gas}},}\ }\href {\doibase 10.1103/PhysRev.136.B864} {\bibfield
  {journal} {\bibinfo  {journal} {Phys. Rev.}\ }\textbf {\bibinfo {volume}
  {136}},\ \bibinfo {pages} {B864--B871} (\bibinfo {year} {1964})}\BibitemShut
  {NoStop}%
\bibitem [{\citenamefont {Kohn}\ and\ \citenamefont {Sham}(1965)}]{Kohn1965}%
  \BibitemOpen
  \bibfield  {author} {\bibinfo {author} {\bibfnamefont {W.}~\bibnamefont
  {Kohn}}\ and\ \bibinfo {author} {\bibfnamefont {L.~J.}\ \bibnamefont
  {Sham}},\ }\bibfield  {title} {\enquote {\bibinfo {title} {{Self-Consistent
  Equations Including Exchange and Correlation Effects}},}\ }\href {\doibase
  10.1103/PhysRev.140.A1133} {\bibfield  {journal} {\bibinfo  {journal} {Phys.
  Rev.}\ }\textbf {\bibinfo {volume} {140}},\ \bibinfo {pages} {A1133--A1138}
  (\bibinfo {year} {1965})}\BibitemShut {NoStop}%
\bibitem [{\citenamefont {Manninen}\ and\ \citenamefont
  {Vaara}(2006)}]{Manninen2006}%
  \BibitemOpen
  \bibfield  {author} {\bibinfo {author} {\bibfnamefont {P.}~\bibnamefont
  {Manninen}}\ and\ \bibinfo {author} {\bibfnamefont {J.}~\bibnamefont
  {Vaara}},\ }\bibfield  {title} {\enquote {\bibinfo {title} {{Systematic
  Gaussian basis-set limit using completeness-optimized primitive sets. A case
  for magnetic properties.}}}\ }\href {\doibase 10.1002/jcc.20358} {\bibfield
  {journal} {\bibinfo  {journal} {J. Comput. Chem.}\ }\textbf {\bibinfo
  {volume} {27}},\ \bibinfo {pages} {434--45} (\bibinfo {year}
  {2006})}\BibitemShut {NoStop}%
\bibitem [{\citenamefont {Lehtola}(2015)}]{Lehtola2015}%
  \BibitemOpen
  \bibfield  {author} {\bibinfo {author} {\bibfnamefont {S.}~\bibnamefont
  {Lehtola}},\ }\bibfield  {title} {\enquote {\bibinfo {title} {{Automatic
  algorithms for completeness-optimization of Gaussian basis sets}},}\ }\href
  {\doibase 10.1002/jcc.23802} {\bibfield  {journal} {\bibinfo  {journal} {J.
  Comput. Chem.}\ }\textbf {\bibinfo {volume} {36}},\ \bibinfo {pages}
  {335--347} (\bibinfo {year} {2015})}\BibitemShut {NoStop}%
\bibitem [{\citenamefont {Lehtola}(2019{\natexlab{f}})}]{Lehtola2019}%
  \BibitemOpen
  \bibfield  {author} {\bibinfo {author} {\bibfnamefont {S.}~\bibnamefont
  {Lehtola}},\ }\bibfield  {title} {\enquote {\bibinfo {title} {{Assessment of
  Initial Guesses for Self-Consistent Field Calculations. Superposition of
  Atomic Potentials: Simple yet Efficient}},}\ }\href {\doibase
  10.1021/acs.jctc.8b01089} {\bibfield  {journal} {\bibinfo  {journal} {J.
  Chem. Theory Comput.}\ }\textbf {\bibinfo {volume} {15}},\ \bibinfo {pages}
  {1593--1604} (\bibinfo {year} {2019}{\natexlab{f}})},\ \Eprint
  {http://arxiv.org/abs/1810.11659} {arXiv:1810.11659} \BibitemShut {NoStop}%
\bibitem [{\citenamefont {Maschio}\ and\ \citenamefont
  {Kirtman}(2019)}]{Maschio2020}%
  \BibitemOpen
  \bibfield  {author} {\bibinfo {author} {\bibfnamefont {L.}~\bibnamefont
  {Maschio}}\ and\ \bibinfo {author} {\bibfnamefont {B.}~\bibnamefont
  {Kirtman}},\ }\bibfield  {title} {\enquote {\bibinfo {title} {{Coupled
  Perturbation Theory Approach to Dual Basis Sets for Molecules and Solids. 1.
  General Theory and Application to Molecules}},}\ }\href {\doibase
  10.1021/acs.jctc.9b00922} {\bibfield  {journal} {\bibinfo  {journal} {J.
  Chem. Theory Comput.}\ ,\ \bibinfo {pages} {acs.jctc.9b00922}} (\bibinfo
  {year} {2019})}\BibitemShut {NoStop}%
\bibitem [{\citenamefont {L{\"{o}}wdin}(1956)}]{Lowdin1956}%
  \BibitemOpen
  \bibfield  {author} {\bibinfo {author} {\bibfnamefont {P.-O.}\ \bibnamefont
  {L{\"{o}}wdin}},\ }\bibfield  {title} {\enquote {\bibinfo {title} {{Quantum
  theory of cohesive properties of solids}},}\ }\href {\doibase
  10.1080/00018735600101155} {\bibfield  {journal} {\bibinfo  {journal} {Adv.
  Phys.}\ }\textbf {\bibinfo {volume} {5}},\ \bibinfo {pages} {1--171}
  (\bibinfo {year} {1956})}\BibitemShut {NoStop}%
\bibitem [{\citenamefont {Jensen}(1999)}]{Jensen1999}%
  \BibitemOpen
  \bibfield  {author} {\bibinfo {author} {\bibfnamefont {F.}~\bibnamefont
  {Jensen}},\ }\bibfield  {title} {\enquote {\bibinfo {title} {{The basis set
  convergence of the Hartree--Fock energy for \ce{H2}}},}\ }\href {\doibase
  10.1063/1.478567} {\bibfield  {journal} {\bibinfo  {journal} {J. Chem.
  Phys.}\ }\textbf {\bibinfo {volume} {110}},\ \bibinfo {pages} {6601}
  (\bibinfo {year} {1999})}\BibitemShut {NoStop}%
\bibitem [{\citenamefont {Bakken}\ and\ \citenamefont
  {Helgaker}(2004)}]{Bakken2004}%
  \BibitemOpen
  \bibfield  {author} {\bibinfo {author} {\bibfnamefont {V.}~\bibnamefont
  {Bakken}}\ and\ \bibinfo {author} {\bibfnamefont {T.}~\bibnamefont
  {Helgaker}},\ }\bibfield  {title} {\enquote {\bibinfo {title} {{The expansion
  of hydrogen states in Gaussian orbitals}},}\ }\href {\doibase
  10.1007/s00214-004-0573-4} {\bibfield  {journal} {\bibinfo  {journal} {Theor.
  Chem. Acc.}\ }\textbf {\bibinfo {volume} {112}},\ \bibinfo {pages} {124--134}
  (\bibinfo {year} {2004})}\BibitemShut {NoStop}%
\bibitem [{\citenamefont {Petersson}\ \emph {et~al.}(2003)\citenamefont
  {Petersson}, \citenamefont {Zhong}, \citenamefont {Montgomery},\ and\
  \citenamefont {Frisch}}]{Petersson2003}%
  \BibitemOpen
  \bibfield  {author} {\bibinfo {author} {\bibfnamefont {G.~A.}\ \bibnamefont
  {Petersson}}, \bibinfo {author} {\bibfnamefont {S.}~\bibnamefont {Zhong}},
  \bibinfo {author} {\bibfnamefont {J.~A.}\ \bibnamefont {Montgomery}}, \ and\
  \bibinfo {author} {\bibfnamefont {M.~J.}\ \bibnamefont {Frisch}},\ }\bibfield
   {title} {\enquote {\bibinfo {title} {{On the optimization of Gaussian basis
  sets}},}\ }\href {\doibase 10.1063/1.1516801} {\bibfield  {journal} {\bibinfo
   {journal} {J. Chem. Phys.}\ }\textbf {\bibinfo {volume} {118}},\ \bibinfo
  {pages} {1101} (\bibinfo {year} {2003})}\BibitemShut {NoStop}%
\bibitem [{\citenamefont {Jensen}(2000)}]{Jensen2000}%
  \BibitemOpen
  \bibfield  {author} {\bibinfo {author} {\bibfnamefont {F.}~\bibnamefont
  {Jensen}},\ }\bibfield  {title} {\enquote {\bibinfo {title} {{The basis set
  convergence of the Hartree--Fock energy for \ce{H3+}, \ce{Li2}, and
  \ce{N2}}},}\ }\href {\doibase 10.1007/s002140000174} {\bibfield  {journal}
  {\bibinfo  {journal} {Theor. Chim. Acta}\ }\textbf {\bibinfo {volume}
  {104}},\ \bibinfo {pages} {484--490} (\bibinfo {year} {2000})}\BibitemShut
  {NoStop}%
\bibitem [{\citenamefont {Raffenetti}(1973)}]{Raffenetti1973a}%
  \BibitemOpen
  \bibfield  {author} {\bibinfo {author} {\bibfnamefont {R.~C.}\ \bibnamefont
  {Raffenetti}},\ }\bibfield  {title} {\enquote {\bibinfo {title}
  {{Even-tempered atomic orbitals. II. Atomic SCF wavefunctions in terms of
  even-tempered exponential bases}},}\ }\href {\doibase 10.1063/1.1679962}
  {\bibfield  {journal} {\bibinfo  {journal} {J. Chem. Phys.}\ }\textbf
  {\bibinfo {volume} {59}},\ \bibinfo {pages} {5936--5949} (\bibinfo {year}
  {1973})}\BibitemShut {NoStop}%
\bibitem [{\citenamefont {Cherkes}, \citenamefont {Klaiman},\ and\
  \citenamefont {Moiseyev}(2009)}]{Cherkes2009}%
  \BibitemOpen
  \bibfield  {author} {\bibinfo {author} {\bibfnamefont {I.}~\bibnamefont
  {Cherkes}}, \bibinfo {author} {\bibfnamefont {S.}~\bibnamefont {Klaiman}}, \
  and\ \bibinfo {author} {\bibfnamefont {N.}~\bibnamefont {Moiseyev}},\
  }\bibfield  {title} {\enquote {\bibinfo {title} {{Spanning the Hilbert space
  with an even tempered Gaussian basis set}},}\ }\href {\doibase
  10.1002/qua.22090} {\bibfield  {journal} {\bibinfo  {journal} {Int. J.
  Quantum Chem.}\ }\textbf {\bibinfo {volume} {109}},\ \bibinfo {pages}
  {2996--3002} (\bibinfo {year} {2009})}\BibitemShut {NoStop}%
\bibitem [{\citenamefont {Feller}\ and\ \citenamefont
  {Ruedenberg}(1979)}]{Feller1979}%
  \BibitemOpen
  \bibfield  {author} {\bibinfo {author} {\bibfnamefont {D.~F.}\ \bibnamefont
  {Feller}}\ and\ \bibinfo {author} {\bibfnamefont {K.}~\bibnamefont
  {Ruedenberg}},\ }\bibfield  {title} {\enquote {\bibinfo {title} {{Systematic
  approach to extended even-tempered orbital bases for atomic and molecular
  calculations}},}\ }\href {\doibase 10.1007/BF00547681} {\bibfield  {journal}
  {\bibinfo  {journal} {Theor. Chim. Acta}\ }\textbf {\bibinfo {volume} {52}},\
  \bibinfo {pages} {231--251} (\bibinfo {year} {1979})}\BibitemShut {NoStop}%
\bibitem [{\citenamefont {Chong}(1995)}]{Chong1995}%
  \BibitemOpen
  \bibfield  {author} {\bibinfo {author} {\bibfnamefont {D.~P.}\ \bibnamefont
  {Chong}},\ }\bibfield  {title} {\enquote {\bibinfo {title} {{Completeness
  profiles of one-electron basis sets}},}\ }\href {\doibase 10.1139/v95-011}
  {\bibfield  {journal} {\bibinfo  {journal} {Can. J. Chem.}\ }\textbf
  {\bibinfo {volume} {73}},\ \bibinfo {pages} {79--83} (\bibinfo {year}
  {1995})}\BibitemShut {NoStop}%
\bibitem [{\citenamefont {Sun}\ \emph {et~al.}(2018)\citenamefont {Sun},
  \citenamefont {Berkelbach}, \citenamefont {Blunt}, \citenamefont {Booth},
  \citenamefont {Guo}, \citenamefont {Li}, \citenamefont {Liu}, \citenamefont
  {McClain}, \citenamefont {Sayfutyarova}, \citenamefont {Sharma},
  \citenamefont {Wouters},\ and\ \citenamefont {Chan}}]{Sun2018}%
  \BibitemOpen
  \bibfield  {author} {\bibinfo {author} {\bibfnamefont {Q.}~\bibnamefont
  {Sun}}, \bibinfo {author} {\bibfnamefont {T.~C.}\ \bibnamefont {Berkelbach}},
  \bibinfo {author} {\bibfnamefont {N.~S.}\ \bibnamefont {Blunt}}, \bibinfo
  {author} {\bibfnamefont {G.~H.}\ \bibnamefont {Booth}}, \bibinfo {author}
  {\bibfnamefont {S.}~\bibnamefont {Guo}}, \bibinfo {author} {\bibfnamefont
  {Z.}~\bibnamefont {Li}}, \bibinfo {author} {\bibfnamefont {J.}~\bibnamefont
  {Liu}}, \bibinfo {author} {\bibfnamefont {J.~D.}\ \bibnamefont {McClain}},
  \bibinfo {author} {\bibfnamefont {E.~R.}\ \bibnamefont {Sayfutyarova}},
  \bibinfo {author} {\bibfnamefont {S.}~\bibnamefont {Sharma}}, \bibinfo
  {author} {\bibfnamefont {S.}~\bibnamefont {Wouters}}, \ and\ \bibinfo
  {author} {\bibfnamefont {G.~K.-L.}\ \bibnamefont {Chan}},\ }\bibfield
  {title} {\enquote {\bibinfo {title} {{PySCF: the Python-based simulations of
  chemistry framework}},}\ }\href {\doibase 10.1002/wcms.1340} {\bibfield
  {journal} {\bibinfo  {journal} {Wiley Interdiscip. Rev. Comput. Mol. Sci.}\
  }\textbf {\bibinfo {volume} {8}},\ \bibinfo {pages} {e1340} (\bibinfo {year}
  {2018})},\ \Eprint {http://arxiv.org/abs/1701.08223} {arXiv:1701.08223}
  \BibitemShut {NoStop}%
\bibitem [{\citenamefont {Lehtola}(2018)}]{erkale}%
  \BibitemOpen
  \bibfield  {author} {\bibinfo {author} {\bibfnamefont {S.}~\bibnamefont
  {Lehtola}},\ }\href {https://github.com/susilehtola/erkale} {\enquote
  {\bibinfo {title} {{ERKALE -- HF/DFT from Hel}},}\ } (\bibinfo {year}
  {2018})\BibitemShut {NoStop}%
\bibitem [{\citenamefont {Lehtola}\ \emph {et~al.}(2012)\citenamefont
  {Lehtola}, \citenamefont {Hakala}, \citenamefont {Sakko},\ and\ \citenamefont
  {H{\"{a}}m{\"{a}}l{\"{a}}inen}}]{Lehtola2012}%
  \BibitemOpen
  \bibfield  {author} {\bibinfo {author} {\bibfnamefont {J.}~\bibnamefont
  {Lehtola}}, \bibinfo {author} {\bibfnamefont {M.}~\bibnamefont {Hakala}},
  \bibinfo {author} {\bibfnamefont {A.}~\bibnamefont {Sakko}}, \ and\ \bibinfo
  {author} {\bibfnamefont {K.}~\bibnamefont {H{\"{a}}m{\"{a}}l{\"{a}}inen}},\
  }\bibfield  {title} {\enquote {\bibinfo {title} {{ERKALE -- A flexible
  program package for X-ray properties of atoms and molecules}},}\ }\href
  {\doibase 10.1002/jcc.22987} {\bibfield  {journal} {\bibinfo  {journal} {J.
  Comput. Chem.}\ }\textbf {\bibinfo {volume} {33}},\ \bibinfo {pages}
  {1572--1585} (\bibinfo {year} {2012})}\BibitemShut {NoStop}%
\bibitem [{\citenamefont {Jensen}\ \emph {et~al.}(2017)\citenamefont {Jensen},
  \citenamefont {Saha}, \citenamefont {Flores-Livas}, \citenamefont {Huhn},
  \citenamefont {Blum}, \citenamefont {Goedecker},\ and\ \citenamefont
  {Frediani}}]{Jensen2017}%
  \BibitemOpen
  \bibfield  {author} {\bibinfo {author} {\bibfnamefont {S.~R.}\ \bibnamefont
  {Jensen}}, \bibinfo {author} {\bibfnamefont {S.}~\bibnamefont {Saha}},
  \bibinfo {author} {\bibfnamefont {J.~A.}\ \bibnamefont {Flores-Livas}},
  \bibinfo {author} {\bibfnamefont {W.}~\bibnamefont {Huhn}}, \bibinfo {author}
  {\bibfnamefont {V.}~\bibnamefont {Blum}}, \bibinfo {author} {\bibfnamefont
  {S.}~\bibnamefont {Goedecker}}, \ and\ \bibinfo {author} {\bibfnamefont
  {L.}~\bibnamefont {Frediani}},\ }\bibfield  {title} {\enquote {\bibinfo
  {title} {{The Elephant in the Room of Density Functional Theory
  Calculations}},}\ }\href {\doibase 10.1021/acs.jpclett.7b00255} {\bibfield
  {journal} {\bibinfo  {journal} {J. Phys. Chem. Lett.}\ }\textbf {\bibinfo
  {volume} {8}},\ \bibinfo {pages} {1449--1457} (\bibinfo {year} {2017})},\
  \Eprint {http://arxiv.org/abs/1702.00957} {arXiv:1702.00957} \BibitemShut
  {NoStop}%
\bibitem [{\citenamefont {Jensen}(2017)}]{Jensen2017b}%
  \BibitemOpen
  \bibfield  {author} {\bibinfo {author} {\bibfnamefont {F.}~\bibnamefont
  {Jensen}},\ }\bibfield  {title} {\enquote {\bibinfo {title} {{How Large is
  the Elephant in the Density Functional Theory Room?}}}\ }\href {\doibase
  10.1021/acs.jpca.7b04760} {\bibfield  {journal} {\bibinfo  {journal} {J.
  Phys. Chem. A}\ }\textbf {\bibinfo {volume} {121}},\ \bibinfo {pages}
  {6104--6107} (\bibinfo {year} {2017})},\ \Eprint
  {http://arxiv.org/abs/1704.08832} {arXiv:1704.08832} \BibitemShut {NoStop}%
\bibitem [{\citenamefont {Feller}\ and\ \citenamefont
  {Dixon}(2018)}]{Feller2018}%
  \BibitemOpen
  \bibfield  {author} {\bibinfo {author} {\bibfnamefont {D.}~\bibnamefont
  {Feller}}\ and\ \bibinfo {author} {\bibfnamefont {D.~A.}\ \bibnamefont
  {Dixon}},\ }\bibfield  {title} {\enquote {\bibinfo {title} {{Density
  Functional Theory and the Basis Set Truncation Problem with Correlation
  Consistent Basis Sets: Elephant in the Room or Mouse in the Closet?}}}\
  }\href {\doibase 10.1021/acs.jpca.8b00392} {\bibfield  {journal} {\bibinfo
  {journal} {J. Phys. Chem. A}\ }\textbf {\bibinfo {volume} {122}},\ \bibinfo
  {pages} {2598--2603} (\bibinfo {year} {2018})}\BibitemShut {NoStop}%
\bibitem [{\citenamefont {Parrish}\ \emph {et~al.}(2017)\citenamefont
  {Parrish}, \citenamefont {Burns}, \citenamefont {Smith}, \citenamefont
  {Simmonett}, \citenamefont {DePrince}, \citenamefont {Hohenstein},
  \citenamefont {Bozkaya}, \citenamefont {Sokolov}, \citenamefont {{Di
  Remigio}}, \citenamefont {Richard}, \citenamefont {Gonthier}, \citenamefont
  {James}, \citenamefont {McAlexander}, \citenamefont {Kumar}, \citenamefont
  {Saitow}, \citenamefont {Wang}, \citenamefont {Pritchard}, \citenamefont
  {Verma}, \citenamefont {Schaefer}, \citenamefont {Patkowski}, \citenamefont
  {King}, \citenamefont {Valeev}, \citenamefont {Evangelista}, \citenamefont
  {Turney}, \citenamefont {Crawford},\ and\ \citenamefont
  {Sherrill}}]{Parrish2017}%
  \BibitemOpen
  \bibfield  {author} {\bibinfo {author} {\bibfnamefont {R.~M.}\ \bibnamefont
  {Parrish}}, \bibinfo {author} {\bibfnamefont {L.~A.}\ \bibnamefont {Burns}},
  \bibinfo {author} {\bibfnamefont {D.~G.~A.}\ \bibnamefont {Smith}}, \bibinfo
  {author} {\bibfnamefont {A.~C.}\ \bibnamefont {Simmonett}}, \bibinfo {author}
  {\bibfnamefont {A.~E.}\ \bibnamefont {DePrince}}, \bibinfo {author}
  {\bibfnamefont {E.~G.}\ \bibnamefont {Hohenstein}}, \bibinfo {author}
  {\bibfnamefont {U.}~\bibnamefont {Bozkaya}}, \bibinfo {author} {\bibfnamefont
  {A.~Y.}\ \bibnamefont {Sokolov}}, \bibinfo {author} {\bibfnamefont
  {R.}~\bibnamefont {{Di Remigio}}}, \bibinfo {author} {\bibfnamefont {R.~M.}\
  \bibnamefont {Richard}}, \bibinfo {author} {\bibfnamefont {J.~F.}\
  \bibnamefont {Gonthier}}, \bibinfo {author} {\bibfnamefont {A.~M.}\
  \bibnamefont {James}}, \bibinfo {author} {\bibfnamefont {H.~R.}\ \bibnamefont
  {McAlexander}}, \bibinfo {author} {\bibfnamefont {A.}~\bibnamefont {Kumar}},
  \bibinfo {author} {\bibfnamefont {M.}~\bibnamefont {Saitow}}, \bibinfo
  {author} {\bibfnamefont {X.}~\bibnamefont {Wang}}, \bibinfo {author}
  {\bibfnamefont {B.~P.}\ \bibnamefont {Pritchard}}, \bibinfo {author}
  {\bibfnamefont {P.}~\bibnamefont {Verma}}, \bibinfo {author} {\bibfnamefont
  {H.~F.}\ \bibnamefont {Schaefer}}, \bibinfo {author} {\bibfnamefont
  {K.}~\bibnamefont {Patkowski}}, \bibinfo {author} {\bibfnamefont {R.~A.}\
  \bibnamefont {King}}, \bibinfo {author} {\bibfnamefont {E.~F.}\ \bibnamefont
  {Valeev}}, \bibinfo {author} {\bibfnamefont {F.~A.}\ \bibnamefont
  {Evangelista}}, \bibinfo {author} {\bibfnamefont {J.~M.}\ \bibnamefont
  {Turney}}, \bibinfo {author} {\bibfnamefont {T.~D.}\ \bibnamefont
  {Crawford}}, \ and\ \bibinfo {author} {\bibfnamefont {C.~D.}\ \bibnamefont
  {Sherrill}},\ }\bibfield  {title} {\enquote {\bibinfo {title} {{Psi4 1.1: An
  Open-Source Electronic Structure Program Emphasizing Automation, Advanced
  Libraries, and Interoperability}},}\ }\href {\doibase
  10.1021/acs.jctc.7b00174} {\bibfield  {journal} {\bibinfo  {journal} {J.
  Chem. Theory Comput.}\ }\textbf {\bibinfo {volume} {13}},\ \bibinfo {pages}
  {3185--3197} (\bibinfo {year} {2017})}\BibitemShut {NoStop}%
\bibitem [{\citenamefont {Perdew}, \citenamefont {Burke},\ and\ \citenamefont
  {Ernzerhof}(1996)}]{Perdew1996}%
  \BibitemOpen
  \bibfield  {author} {\bibinfo {author} {\bibfnamefont {J.~P.}\ \bibnamefont
  {Perdew}}, \bibinfo {author} {\bibfnamefont {K.}~\bibnamefont {Burke}}, \
  and\ \bibinfo {author} {\bibfnamefont {M.}~\bibnamefont {Ernzerhof}},\
  }\bibfield  {title} {\enquote {\bibinfo {title} {{Generalized Gradient
  Approximation Made Simple}},}\ }\href {\doibase 10.1103/PhysRevLett.77.3865}
  {\bibfield  {journal} {\bibinfo  {journal} {Phys. Rev. Lett.}\ }\textbf
  {\bibinfo {volume} {77}},\ \bibinfo {pages} {3865--3868} (\bibinfo {year}
  {1996})}\BibitemShut {NoStop}%
\bibitem [{\citenamefont {Perdew}, \citenamefont {Burke},\ and\ \citenamefont
  {Ernzerhof}(1997)}]{Perdew1997}%
  \BibitemOpen
  \bibfield  {author} {\bibinfo {author} {\bibfnamefont {J.~P.}\ \bibnamefont
  {Perdew}}, \bibinfo {author} {\bibfnamefont {K.}~\bibnamefont {Burke}}, \
  and\ \bibinfo {author} {\bibfnamefont {M.}~\bibnamefont {Ernzerhof}},\
  }\bibfield  {title} {\enquote {\bibinfo {title} {{Generalized Gradient
  Approximation Made Simple [Phys. Rev. Lett. 77, 3865 (1996)]}},}\ }\href
  {\doibase 10.1103/PhysRevLett.78.1396} {\bibfield  {journal} {\bibinfo
  {journal} {Phys. Rev. Lett.}\ }\textbf {\bibinfo {volume} {78}},\ \bibinfo
  {pages} {1396--1396} (\bibinfo {year} {1997})}\BibitemShut {NoStop}%
\bibitem [{\citenamefont {Hill}\ and\ \citenamefont
  {Wheeler}(1953)}]{Hill1953}%
  \BibitemOpen
  \bibfield  {author} {\bibinfo {author} {\bibfnamefont {D.~L.}\ \bibnamefont
  {Hill}}\ and\ \bibinfo {author} {\bibfnamefont {J.~A.}\ \bibnamefont
  {Wheeler}},\ }\bibfield  {title} {\enquote {\bibinfo {title} {{Nuclear
  Constitution and the Interpretation of Fission Phenomena}},}\ }\href
  {\doibase 10.1103/PhysRev.89.1102} {\bibfield  {journal} {\bibinfo  {journal}
  {Phys. Rev.}\ }\textbf {\bibinfo {volume} {89}},\ \bibinfo {pages}
  {1102--1145} (\bibinfo {year} {1953})}\BibitemShut {NoStop}%
\bibitem [{\citenamefont {Griffin}\ and\ \citenamefont
  {Wheeler}(1957)}]{Griffin1957}%
  \BibitemOpen
  \bibfield  {author} {\bibinfo {author} {\bibfnamefont {J.~J.}\ \bibnamefont
  {Griffin}}\ and\ \bibinfo {author} {\bibfnamefont {J.~A.}\ \bibnamefont
  {Wheeler}},\ }\bibfield  {title} {\enquote {\bibinfo {title} {{Collective
  Motions in Nuclei by the Method of Generator Coordinates}},}\ }\href
  {\doibase 10.1103/PhysRev.108.311} {\bibfield  {journal} {\bibinfo  {journal}
  {Phys. Rev.}\ }\textbf {\bibinfo {volume} {108}},\ \bibinfo {pages}
  {311--327} (\bibinfo {year} {1957})}\BibitemShut {NoStop}%
\bibitem [{\citenamefont {Mohallem}, \citenamefont {Dreizler},\ and\
  \citenamefont {Trsic}(1986)}]{Mohalem1986}%
  \BibitemOpen
  \bibfield  {author} {\bibinfo {author} {\bibfnamefont {J.~R.}\ \bibnamefont
  {Mohallem}}, \bibinfo {author} {\bibfnamefont {R.~M.}\ \bibnamefont
  {Dreizler}}, \ and\ \bibinfo {author} {\bibfnamefont {M.}~\bibnamefont
  {Trsic}},\ }\bibfield  {title} {\enquote {\bibinfo {title} {{A
  Griffin--Hill--Wheeler version of the Hartree--Fock equations}},}\ }\href
  {\doibase 10.1002/qua.560300707} {\bibfield  {journal} {\bibinfo  {journal}
  {Int. J. Quantum Chem.}\ }\textbf {\bibinfo {volume} {30}},\ \bibinfo {pages}
  {45--55} (\bibinfo {year} {1986})}\BibitemShut {NoStop}%
\bibitem [{\citenamefont {Somarjai}(1968)}]{Somarjai1968}%
  \BibitemOpen
  \bibfield  {author} {\bibinfo {author} {\bibfnamefont {R.~L.}\ \bibnamefont
  {Somarjai}},\ }\bibfield  {title} {\enquote {\bibinfo {title} {{Integral
  transform functions. A new class of approximate wave functions}},}\ }\href
  {\doibase 10.1016/0009-2614(68)80037-5} {\bibfield  {journal} {\bibinfo
  {journal} {Chem. Phys. Lett.}\ }\textbf {\bibinfo {volume} {2}},\ \bibinfo
  {pages} {399--401} (\bibinfo {year} {1968})}\BibitemShut {NoStop}%
\bibitem [{\citenamefont {Matsuoka}\ and\ \citenamefont
  {Huzinaga}(1987)}]{Matsuoka1987}%
  \BibitemOpen
  \bibfield  {author} {\bibinfo {author} {\bibfnamefont {O.}~\bibnamefont
  {Matsuoka}}\ and\ \bibinfo {author} {\bibfnamefont {S.}~\bibnamefont
  {Huzinaga}},\ }\bibfield  {title} {\enquote {\bibinfo {title} {{Relativistic
  well-tempered gaussian basis sets}},}\ }\href {\doibase
  10.1016/0009-2614(87)80488-8} {\bibfield  {journal} {\bibinfo  {journal}
  {Chem. Phys. Lett.}\ }\textbf {\bibinfo {volume} {140}},\ \bibinfo {pages}
  {567--571} (\bibinfo {year} {1987})}\BibitemShut {NoStop}%
\bibitem [{\citenamefont {Dyall}\ and\ \citenamefont
  {F{\ae}gri}(1996)}]{Dyall1996}%
  \BibitemOpen
  \bibfield  {author} {\bibinfo {author} {\bibfnamefont {K.~G.}\ \bibnamefont
  {Dyall}}\ and\ \bibinfo {author} {\bibfnamefont {K.}~\bibnamefont
  {F{\ae}gri}},\ }\bibfield  {title} {\enquote {\bibinfo {title} {{Optimization
  of Gaussian basis sets for Dirac--Hartree--Fock calculations}},}\ }\href
  {\doibase 10.1007/BF00190154} {\bibfield  {journal} {\bibinfo  {journal}
  {Theor. Chim. Acta}\ }\textbf {\bibinfo {volume} {94}},\ \bibinfo {pages}
  {39--51} (\bibinfo {year} {1996})}\BibitemShut {NoStop}%
\bibitem [{\citenamefont {Ishikawa}, \citenamefont {Baretty},\ and\
  \citenamefont {Binning}(1985)}]{Ishikawa1985}%
  \BibitemOpen
  \bibfield  {author} {\bibinfo {author} {\bibfnamefont {Y.}~\bibnamefont
  {Ishikawa}}, \bibinfo {author} {\bibfnamefont {R.}~\bibnamefont {Baretty}}, \
  and\ \bibinfo {author} {\bibfnamefont {R.~C.}\ \bibnamefont {Binning}},\
  }\bibfield  {title} {\enquote {\bibinfo {title} {{Relativistic Gaussian basis
  set calculations on one-electron ions with a nucleus of finite extent}},}\
  }\href {\doibase 10.1016/0009-2614(85)87169-4} {\bibfield  {journal}
  {\bibinfo  {journal} {Chem. Phys. Lett.}\ }\textbf {\bibinfo {volume}
  {121}},\ \bibinfo {pages} {130--133} (\bibinfo {year} {1985})}\BibitemShut
  {NoStop}%
\bibitem [{\citenamefont {Pyykk{\"{o}}}(2011)}]{Pyykko2011}%
  \BibitemOpen
  \bibfield  {author} {\bibinfo {author} {\bibfnamefont {P.}~\bibnamefont
  {Pyykk{\"{o}}}},\ }\bibfield  {title} {\enquote {\bibinfo {title} {{A
  suggested periodic table up to $Z \leq 172$, based on Dirac--Fock
  calculations on atoms and ions}},}\ }\href {\doibase 10.1039/C0CP01575J}
  {\bibfield  {journal} {\bibinfo  {journal} {Phys. Chem. Chem. Phys.}\
  }\textbf {\bibinfo {volume} {13}},\ \bibinfo {pages} {161--168} (\bibinfo
  {year} {2011})}\BibitemShut {NoStop}%
\end{thebibliography}%
